\documentclass[%
 reprint,
 superscriptaddress,
 amsmath,amssymb,
 aps,
floatfix,
]{revtex4-2}

\usepackage{graphicx}
\usepackage{float} 
\usepackage{dcolumn}
\usepackage{bm}
\usepackage{comment}
\usepackage[caption=false]{subfig}
\usepackage{blindtext}
\usepackage[dvipsnames]{xcolor}
\usepackage[table]{xcolor}

\begin{document}\href{}{}

\title{Effect of glass stability on the low frequency vibrations of vapor deposited glasses}

\author{I. Festi}
 \email{irene.festi@unitn.it}
 \affiliation{Department of Physics, University of Trento, I-38123, Povo, Trento, Italy.}
\author{E. Alfinelli}
 \affiliation{Department of Physics, University of Trento, I-38123, Povo, Trento, Italy.}
\author{D. Bessas}
 \affiliation{European Synchrotron Radiation Facility, BP 220, F-38043 Grenoble, France.}
\author{F. Caporaletti}
 \affiliation{Laboratory  of  Polymer  and  Soft  Matter  Dynamics,  Experimental  Soft  Matter  and  Thermal Physics (EST), Université libre de Bruxelles (ULB), Brussels 1050, Belgium.}
\author{A. I. Chumakov}
 \affiliation{European Synchrotron Radiation Facility, BP 220, F-38043 Grenoble, France.}
\author{M. Moratalla}
 \affiliation{Departamento de Física de la Materia Condensada, Condensed Matter Physics Center (IFIMAC), and Instituto Nicolás Cabrera (INC), Universidad Autónoma de Madrid, 28049 Madrid, Spain.}
\author{M. A. Ramos}
 \affiliation{Departamento de Física de la Materia Condensada, Condensed Matter Physics Center (IFIMAC), and Instituto Nicolás Cabrera (INC), Universidad Autónoma de Madrid, 28049 Madrid, Spain.}
\author{M. Rodríguez-López}
 \affiliation{Departamento de Física, Facultat de Ciències, Universitat Autònoma de Barcelona, 08193 Bellaterra, Spain.}
 \affiliation{Catalan Institute of Nanoscience and Nanotechnology (ICN2), CSIC and BIST, Campus UAB, Bellaterra, 08193 Barcelona, Spain.}
\author{C. Rodríguez-Tinoco}
 \affiliation{Departamento de Física, Facultat de Ciències, Universitat Autònoma de Barcelona, 08193 Bellaterra, Spain.}
 \affiliation{Catalan Institute of Nanoscience and Nanotechnology (ICN2), CSIC and BIST, Campus UAB, Bellaterra, 08193 Barcelona, Spain.}
\author{J. Rodríguez-Viejo}
 \affiliation{Departamento de Física, Facultat de Ciències, Universitat Autònoma de Barcelona, 08193 Bellaterra, Spain.}
 \affiliation{Catalan Institute of Nanoscience and Nanotechnology (ICN2), CSIC and BIST, Campus UAB, Bellaterra, 08193 Barcelona, Spain.}
\author{G. Baldi.}
\email{giacomo.baldi@unitn.it}
\affiliation{Department of Physics, University of Trento, I-38123, Povo, Trento, Italy.}
\date{\today}

\begin{abstract}
 Ultra-stable glasses prepared from the physical vapor deposition of organic molecules present a very low density of two-level states, the kind of glass defects that determine their peculiar low temperature thermal properties. Numerical simulations suggest that quasi-localized harmonic vibrational modes emerge in the soft regions associated with two-level states. However, the connection between the low frequency vibrational modes and the local structural instabilities of glasses remains unexplained. Here we exploit a recently developed spectrograph for nuclear resonant analysis of inelastic X-ray scattering to probe the density of vibrational states of amorphous thin films of ultra-stable and conventional glasses down to an exceptionally low frequency of $\sim 70$ GHz. We show that the glass stability does not affect the harmonic vibrational modes at the lowest frequencies, despite a reduction of almost an order of magnitude in the density of two-level states. At the same time, the vibrational modes at higher frequencies, around the boson peak maximum, are extremely sensitive to the glass stability.  Although we cannot exclude the possible existence of quasi-localized modes in glasses, we show that their presence is not strictly necessary to describe the measured density of low frequency vibrations. The experimental developments here presented pave the way to the solution to the long-standing debate on the low frequency vibrations in glasses.
\end{abstract}

\maketitle

\section{Introduction}
A glass is formed when a liquid is cooled at a sufficiently high cooling rate to avoid crystallization. The solidification process accompanying the transition from the supercooled liquid to the glass can be conveniently described as an evolution of the system within the potential energy landscape~\cite{Debenedetti2001}. The primary structural relaxation process becomes arrested at the glass transition temperature, $T_g$, as the system remains trapped in a local energy minimum. Secondary relaxation processes remain active at lower temperatures, as the glass can explore several almost equivalent configurations corresponding to local minima separated by small energy barriers. As a consequence, the low temperature thermal properties of glasses are markedly different from those of perfect crystalline solids~\cite{Zeller1971} and deviate from the predictions of the Debye's model. The linear temperature dependence of the specific heat, $C_p(T)$, and the quadratic rise with temperature of the thermal conductivity are both satisfactorily explained by the phenomenological tunnelling model, based on the assumption that atoms or groups of atoms tunnel between two-level states (TLS), two closely spaced minima of the potential energy landscape~\cite{Phillips1987}.

An alternative route to glass formation is obtained through the layer by layer deposition of the chosen material on a substrate, resulting in amorphous thin films with relevant technological applications in various fields, such as in the semiconductor industry~\cite{Chan2004,Song2020} or as coatings for optical components~\cite{Addie2022,Abernathy2018}. The physical vapor deposition of specific organic molecules on a substrate held at a fraction of the glass transition temperature, typically 10 - 15\% below $T_g$, has allowed the preparation of amorphous solids with enhanced thermodynamic and kinetic stability, high densities, and exceptional mechanical properties~\cite{Swallen2007,Rodriguez-Tinoco2022}. Recent experiments have shown that ultra-stable glasses obtained in this way present an extraordinarily low density of two level states or even their complete absence~\cite{Perez-Castaneda2014,Moratalla2023}, confirming that they are trapped in deep minima of the energy landscape and suggesting them as candidates of the ``ideal" glass, a glass without defects~\cite{Berthier_2016}.

A universal feature observed in all kinds of amorphous solids is the presence of a peak in $C_p(T)/T^3$, usually termed \textit{boson peak} (BP), located in the same temperature range, around 10 K, where the thermal conductivity shows a plateau. The BP appears as an excess above the Debye level in the density of vibrational states (DOS) at frequencies of about one terahertz \cite{Baldibook}. The nature of the vibrational modes at the BP frequency and below is not understood and several theories for the terahertz vibrations of glasses have been proposed~\cite{Karpov1983,Buchenau1991,vanHove,Grigera2003,Parshin2007,HET_ruocco,Ruffle2008,Chumakov2014,DeGiuli2014,Baggioli2019,Gonzalez-Jimenez2023,Vogel2023,Jiang2024}. Some models assume that an intimate connection is present between the low frequency vibrations and the glass instability associated with TLS~\cite{Karpov1983,Buchenau1991,Parshin2007,Ruffle2008,Baggioli2019,Jiang2024}, while others treat the two as independent physical processes~\cite{vanHove,Grigera2003,HET_ruocco,Chumakov2014,DeGiuli2014,Hu2022,Vogel2023}. 

Recent improvements in numerical simulations have allowed the identification of TLS in model glasses and their study as a function of glass stability, confirming the reduction of TLS in highly stable glasses and supporting the idea of a close connection between TLS and the low frequency vibrations~\cite{Wang_2019,Khomenko_deplet,Khomenko_2021,Ciarella_2023,Xu2024}. The numerical works identify the presence of a peculiar class of vibrational modes, localized around soft elastic spots in the glass and termed ``quasi-localized modes" (QLM), whose density grows as the fourth power of frequency~\cite{Lerner,Angelani,Mizuno,Shcheblanov2020,Bouchbinder2020,Bonfanti2020,Das2021,Bouchbinder2023}, suggesting that they are responsible for the excess of modes at the BP. However, QLM in numerical simulations are observed on large systems with idealized, typically hard-sphere-like, interatomic potentials~\cite{Mizuno,Wang_2019,Khomenko_2021} or in small systems where they can be detected because the reduced system size suppresses the phonon modes~\cite{Lerner,Bouchbinder2020,Bouchbinder2023,Schirmacher2024}. The experimental detection of QLM is difficult because their density is expected to vanish faster than the density of the sound waves as the frequency is lowered, and because the measurement of the DOS is very challenging in the relevant frequency range, between 10 GHz and  1 THz.
Two main questions thus remain open on the nature of the low frequency vibrations in glasses: \emph{i)} are QLM present in real glasses or do they appear in simulations because of finite size effects or exceedingly fast cooling rates? \emph{ii)} is there a connection between TLS and the low frequency vibrations? 

Here we show that an increased glass stability strongly suppresses the TLS density but does not appreciably affect the vibrational modes that lie well below the boson peak frequency. We observe a BP feature in the DOS of an ultra-stable organic molecular glass and find it shifted to slightly higher frequencies and with a significant reduction of intensity compared to the corresponding ordinary glass. Our results show that the vibrational modes at the BP are extremely sensitive to glass stability.  At the same time, the lower frequency harmonic modes are not appreciably influenced by the thermodynamic stability. This result is in apparent contrast with the numerical works based on the swap Monte Carlo algorithm, where both the density of QLM and the sound attenuation are markedly affected by the glass stability~\cite{Wang_2019,Flenner2025,flenner202504}.

We exploit an innovative X-ray spectrometer~\cite{PhysRevLett.123.097402} to probe the DOS of an ultra-stable glass and of the corresponding ordinary glass down to $\sim 70$ GHz, opening a frequency window of more than one decade below the BP maximum. 
We determine the DOS in absolute units and find quantitative agreement with previously measured specific heat data~\cite{Moratalla2023}. The comparison allows us to precisely estimate the density of TLS, the Debye level, and to quantify the departure of the DOS from the Debye prediction. We then compare the measured DOS with two models for the vibrations in glasses, the soft potential model~\cite{Karpov1983,Buchenau1991,Parshin2007} and the heterogeneous elasticity theory~\cite{HET_ruocco,articlemodel,SCHIRMACHER2015133,PhysRevB.104.134106}, and show that both theories are unable to provide a satisfactory explanation for the boson peak reduction with increased stability. We conclude the paper discussing the relevance of sound attenuation as the main mechanism giving rise to a deviation of the DOS from the Debye law. This explains the apparent contradiction with the numerical simulations, since the QLM density is very small in the studied samples.

\section{Experimental methods in brief}
Experimental information on the low frequency vibrational modes of glasses is traditionally obtained by means of Raman~\cite{Jackle1981} or neutron scattering~\cite{Buchenau1984} spectroscopies. The Raman scattering cross section is influenced by the molecular polarizability at optical frequencies and does not give direct access to the density of vibrational states. A quantity similar to the real DOS can be measured in samples that scatter neutrons coherently, with inelastic neutron scattering and the incoherent approximation, which is fulfilled only when the signal is averaged over a large range of exchanged wavevectors. This requirement severely constrains the energy resolution, limiting the minimum measurable frequency to $\sim 300$ GHz~\cite{Fabiani2008}. Higher resolutions can be reached in samples that scatter neutrons incoherently, typically those containing hydrogen~\cite{Monnier2021}. However, neutron scattering methods require the use of bulk samples to obtain a sufficient signal-to-noise ratio and cannot be easily applied to the measurement of the DOS of a thin film. Inelastic scattering of X-rays is an alternative method that can be employed on smaller samples, thanks to focal spots that can be as small as a few microns. X-ray scattering spectrometers are usually limited in their energy resolution because the X-ray energy is much higher than that of neutrons of the same wavelength~\cite{PhysRevLett.112.025502}. 

\begin{figure*}[ht]
    \centering
    \includegraphics[width=\linewidth]{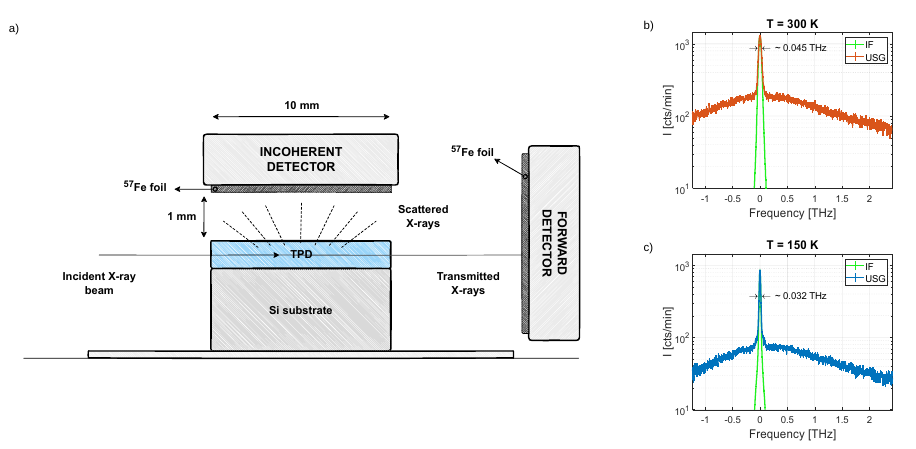}
    \caption{Inelastic X-ray scattering with nuclear resonance analysis. a): simplified sketch of the experimental setup at the sample stage (the drawing is not to scale)~\cite{article99}. b)-c): The raw data measured on the USG sample at 300 K (b)) and at 150 K (c)). The red and blue data represent the scattered intensity collected by the detector located above the sample, while the green data correspond to the instrumental function measured by the forward detector. The intensity scale refers to the sample signal.}
    \label{Figs: raw}
\end{figure*}
Here we show that a recently developed hard X-ray spectrograph, based on a three crystals setup, can be successfully applied to measure the DOS of amorphous thin films with an unprecedented frequency resolution of $\sim 30$ GHz ($\sim 130 \mu$eV), improving by more than an order of magnitude the performance of previous spectrometers. The technique is based on the nuclear resonance analysis of inelastic X-ray scattering~\cite{Chumakov1996}. The monochromatic X-ray beam impinges laterally on the thin amorphous film, as shown in the simplified sketch of the sample stage in Figure~\ref{Figs: raw}a). The sample, a glass of an organic molecule of TPD, is partially transparent to X-rays so that the transmitted and the scattered photons can be collected simultaneously. A detector placed in the forward direction measures the instrumental response function, while the scattered signal is collected over a broad solid angle [0.7 - 14.6] \AA$^{-1}$ by a detector positioned 1 mm above the sample. The transmitted and the scattered intensities are analyzed by means of the nuclear resonance of $\alpha-^{57}$Fe, which provides a bandwidth in detection of $\sim 0.5 \mu$eV and allows us to fully exploit the narrow bandwidth of the incident beam.

We measured an ultra-stable (USG) and an ordinary (OG) glass of TPD. Vapor deposited ultra-stable glasses of indomethacin~\cite{Perez-Castaneda2014} and of TPD~\cite{Moratalla2023} are known to have a very low density of TLS. We chose TPD for our study because it is more resistant to X-ray radiation than indomethacin. Both the USG and OG glasses are measured at room temperature and an additional measurement at 150 K is performed on the USG. The signal collected on the USG at room temperature, for approximately 30 hours, is reported in panel b) of Figure~\ref{Figs: raw}, while panel c) shows the signal at 150 K of an equivalent sample, after more than three days of integration. The figures include the instrumental resolution functions for both measurements (green lines) and the frequency resolution (full width at half maximum, FWHM), which was slightly improved in the low temperature measurement, performed during a second beam-time. The intensity is plotted in a log scale to highlight the inelastic signal over the intense elastic line due to the structural disorder of the glass. More details on the sample preparation and on the experimental technique are given in appendix~\ref{Methods}. 

\section{The reduced density of vibrational states}
\begin{figure}[ht]
    \includegraphics[width=0.95\linewidth]{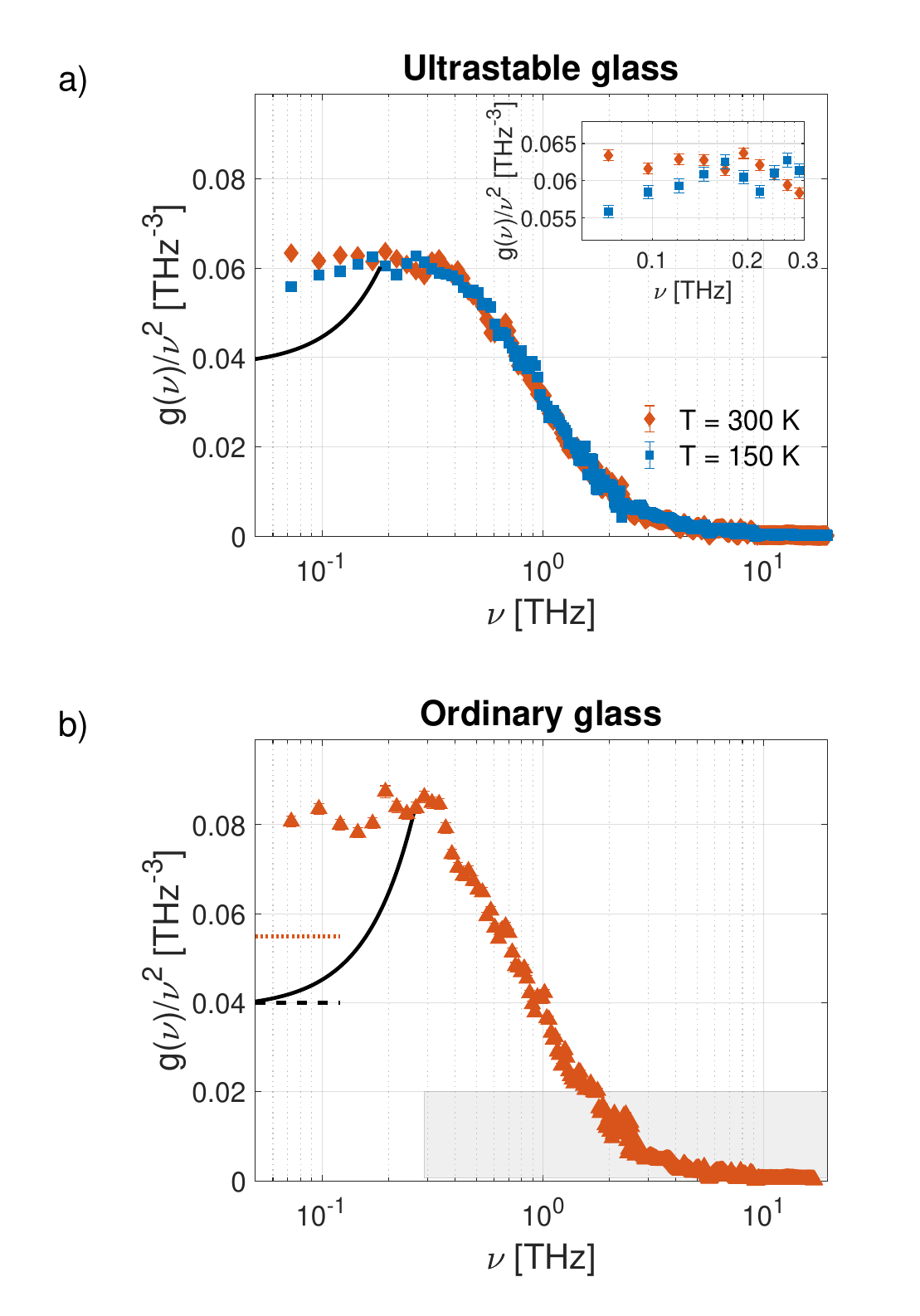}
    \caption{The reduced density of vibrational states. a): ultra-stable glass measured at 300 K (red diamonds) and at 150 K (blue squares); b): ordinary glass at 300 K (red triangles). The horizontal lines in panel b) are the Debye levels estimated from the elastic moduli at 300 K (red dotted), and at 30 K (black dashed). The inset of panel a) highlights the low frequency region, where a difference between the two temperatures is visible below $\sim 0.15$ THz. The parabolic black lines are an estimate of the low frequency part of the harmonic DOS, based on eq.~\eqref{Eq: par}. The grey rectangle (panel b)) represents the energy region accessible with standard X-ray monochromators.}
  \label{Figs: dos}
\end{figure}

The DOS in absolute units is computed from the raw spectra after subtraction of the elastic line and evaluation of the multi-phonon contribution, as detailed in appendix~\ref{Methods: data analysis}. 
The obtained reduced density of states, $g(\nu)/\nu^2$, is shown in Figure~\ref{Figs: dos}a) for the USG and in Figure~\ref{Figs: dos}b) for the OG. The abscissae are plotted in log scale to highlight the low frequency range, where the measurement is reliable above $\sim 0.07$ THz, approximately twice the full width of the instrumental resolution.
A faint boson peak is visible at $\sim 0.3$ THz at low temperatures in the USG. The comparison of the OG and USG allows us to reveal that the BP is extremely sensitive to glass stability, its intensity being significantly suppressed in the USG, in agreement with the specific heat data~\cite{Moratalla2023}. 

The measured reduced DOS for both glasses at room temperature are almost flat between 0.07 and 0.3 THz and are $\sim 50$\% higher than the corresponding Debye levels determined from the elastic moduli~\cite{Moratalla2023}, plotted as horizontal lines in Figure~\ref{Figs: dos}b) at two temperatures (30 K and 300 K). A variation of the reduced DOS of the USG with temperature is visible below $\sim 0.15$ THz, as highlighted in the inset of the upper panel, indicating that the harmonic approximation is valid only at higher frequencies or at lower temperatures. Without loss of generality, we can write the DOS of a glass as the sum of a relaxational, $g_{rel}(\nu,T)$, and a vibrational, $g_{vib}(\nu,T)$, component:
\begin{equation}\label{eq:relvib}
    g(\nu,T) = g_{rel}(\nu,T) + g_{vib}(\nu,T).
\end{equation}
The measured quantity is the total DOS, $g(\nu,T)$, and both its components can be temperature dependent. The relaxational part is normally attributed to thermally activated relaxation processes within double well potentials~\cite{Buchenau88,Buchenau2007,Caponi2007,Ruffle2008} and becomes negligible at temperatures where the BP appears in $C_p/T^3$, around 10 K. Anharmonic effects can give rise to a temperature dependence of the vibrational component of the DOS~\cite{Baldi2014}, as reported also for crystals~\cite{Ren2023}. 
At low temperatures, where both relaxational and anharmonic effects can be neglected, we can write the DOS at sufficiently low frequencies, well below the BP maximum, as:
\begin{equation}\label{Eq: par}
      \lim_{T \to 0 K} g(\nu,T) = g_{vib}(\nu,0) \sim \frac{3}{\nu_D^3}\nu^2 + A_{ex} \nu^4,
\end{equation}
with $\nu_D$ the Debye frequency and $A_{ex}$ the coefficient of the first departure from the Debye prediction, where ``ex" stands for ``excess". Various processes can contribute to the quartic term: i) the presence of quasi localized modes with a density $g_{QLM} \propto \nu^4$~\cite{Buchenaubook,Bouchbinder2020,Xu2024}; ii) the Rayleigh scattering of sound, which gives rise to a $\nu^4$ dependence of the sound attenuation and to a $\propto \nu^4$ contribution to $g(\nu)$~\cite{HET_ruocco,Baldi2014}; iii) the first departure from the Debye level is $\propto \nu^4$ also in crystals, when approaching the first van Hove singularity.

The continuous lines in Figure~\ref{Figs: dos} represent our best estimate of the reduced DOS at low frequency in the low temperature limit, obtained by a proper analysis of the specific heat data, as detailed in the next section.

\section{Low temperature specific heat}
As seen from Figure~\ref{Figs: dos}, the low-energy part of the reduced DOS lies above the Debye level, indicating the presence of an-harmonic and/or relaxational effects. These contributions can be separated from the harmonic part of the DOS using their different effect on the specific heat. Namely, the Debye frequency, $\nu_D$, and the first departure of $g(\nu,0)$ from the Debye prediction can be estimated from the measured DOS and the low temperature specific heat. In describing the specific heat one has to take into account the contribution from tunneling in TLS, which gives rise to a term proportional to temperature.
The specific heat per unit mass can be evaluated as~\cite{Ramosbook}:
\begin{equation}\label{E:Cp}
      C_p(T) =  \frac{3}{M_{at}} \int_0^{\infty} d \nu g(\nu,0) \frac{\partial n(T,\nu)}{\partial T} h \nu + C_{TLS} T,
\end{equation}
where $M_{at}$ is the average atomic mass and $n(T,\nu) = \left[ \exp(h\nu/k_BT)-1 \right]^{-1}$ is the Bose population factor for phonons, with $k_B$ the Boltzmann constant and $h$ the Planck constant. The first term on the right-hand side of the equation is the normal modes contribution to $C_p$, expressed in terms of the harmonic DOS, with area normalized to unity. The contribution from two level states is described by the second term, where we assume a frequency-independent density (number per unit volume and unit frequency) of TLS: $n_{TLS}$. The relation between $n_{TLS}$ and the coefficient $C_{TLS}$ within the standard tunneling model is reported in eq.~\eqref{eq:nTLS} in appendix~\ref{App:Cp}.

\begin{figure}[ht]
    \centering
\includegraphics[width=0.95\linewidth]{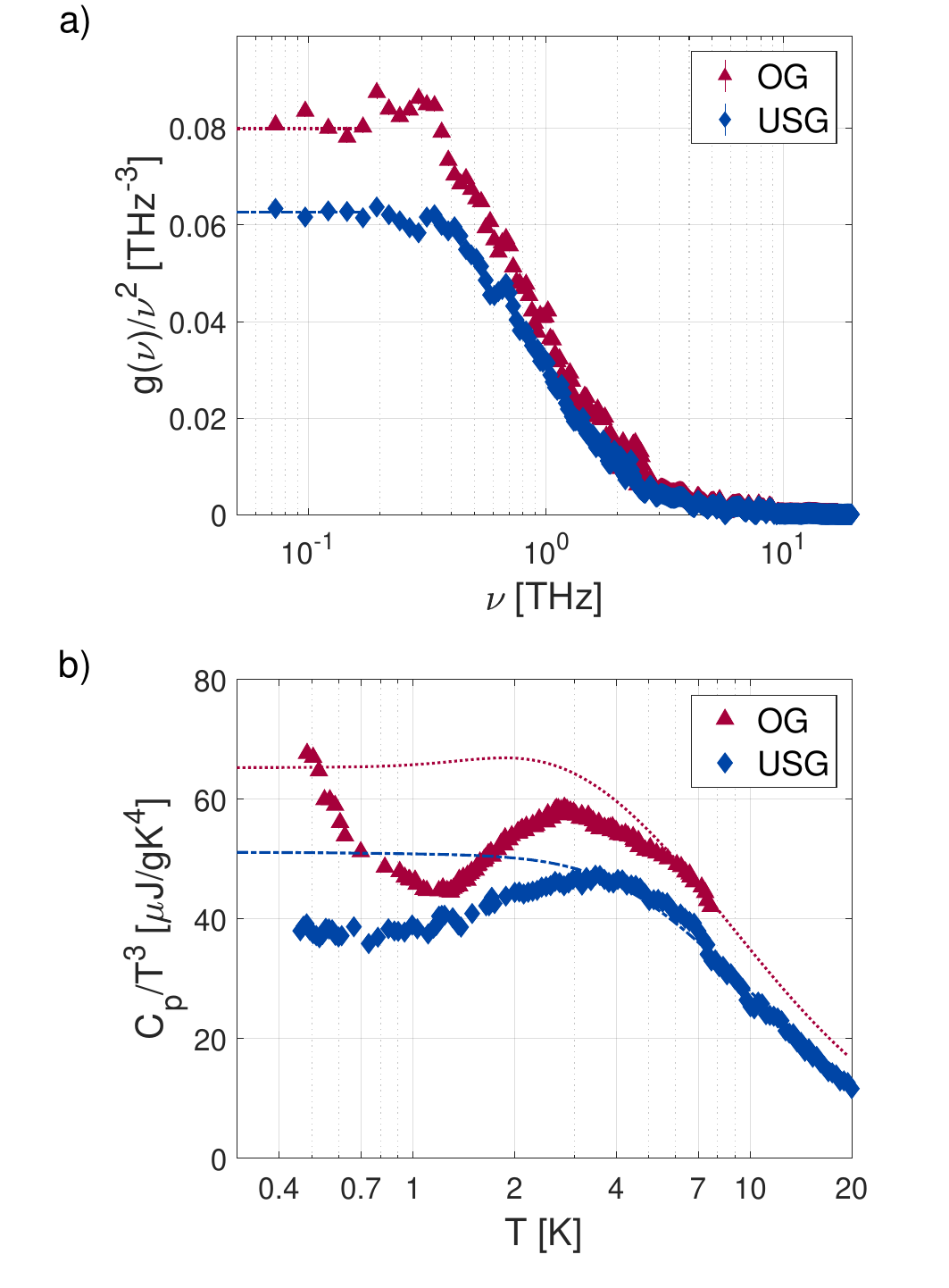}
\caption{Specific heat calculation assuming the measured DOS to be harmonic. Panel a): reduced DOS of the OG (red triangles) and USG (blue diamonds) at room temperature. In this first attempt we treat the DOS as temperature independent and we consider a constant value at low frequencies (lines). Panel b): specific heat divided by the cube of temperature of the OG and USG. The lines are computed using eq.~\eqref{eq:Cp} with the measured room temperature DOS and the low frequency extrapolation shown in a). The curves do not accurately describe $C_p/T^3$, because the DOS is affected by a temperature dependence at low frequencies and the specific heat reflects the DOS at low temperatures, much lower than 300 K, as discussed in the text.} \label{Figs/Analysis: first}
\end{figure}
Let's first consider the room temperature DOS of the USG and OG glasses. We start by calculating $C_p$ assuming a constant reduced DOS at frequencies below the BP, as illustrated in Figure~\ref{Figs/Analysis: first}a), and ignoring the TLS contribution. The calculated $C_p$ are shown in Figure~\ref{Figs/Analysis: first}b) as dotted and dash-dotted lines and specific heat data are reported as diamonds and triangles.
The result does not reproduce the low temperature behaviour of $C_p$, clarifying that we cannot neglect the temperature dependence of the DOS and the difference between the low frequency values of $g(\nu,T)$ and the Debye levels, as done improperly in recent papers~\cite{Moriel2024a, Moriel2024b}. The specific heat data reflect the sample dynamics at low temperatures ($T < 10$ K), where the DOS can be well described by eq.~\eqref{Eq: par}, while the measured DOS at low frequency has a value significantly higher than the Debye level, as shown in Figure~\ref{Figs: dos}. An estimate of the low temperature DOS and the inclusion of the TLS contribution is necessary for a quantitative description of the measured specific heat. 

\begin{figure}[h]
  \centering
    \includegraphics[width=0.95\linewidth]{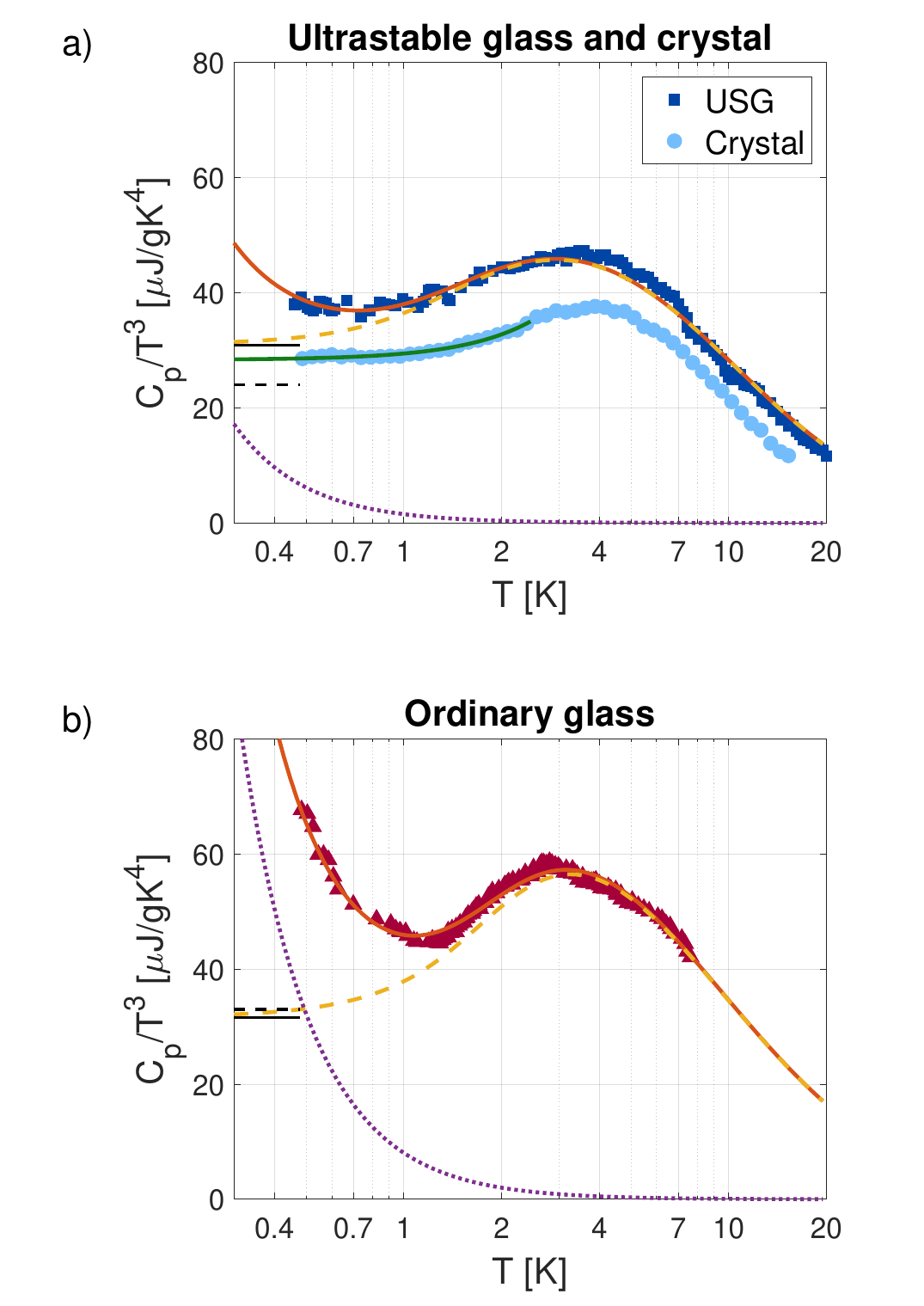}
    \caption{The specific heat over $T^3$. The calculated $C_p$ (red solid line) compared to the experimental ones for the USG (blue squares) in a) and for the OG (red triangles) in b). The harmonic contribution (yellow dashed line) and the TLS contribution (violet dotted line) are shown as well. The Debye level found with the fitting procedure is plotted as a black horizontal line. A significant discrepancy with the $T \to 0$ K estimate from the elastic moduli (black dashed line) is found for the USG, while the two values coincide for the OG. The specific heat of the crystal is also shown for comparison (light-blue circles) in panel a) together with the fit to a parabola at low temperatures (green line), as discussed in the text. The specific heat data are from ref~\cite{Moratalla2023}.}
    \label{Figs: Cp}
\end{figure}

\section{Estimate of the harmonic DOS}\label{S:harm}
Although the density of states measured with our X-ray spectrograph are affected by an-harmonic effects at low frequencies, we can determine a good approximation of the harmonic DOS exploiting the information included in the low temperature specific heat data.
We employ the following procedure: \emph{i)} we model $g(\nu,0)$ using eq.~\eqref{Eq: par} below a frequency $\nu_0$ and the measured DOS above. The parameter $\nu_0$ is close to the BP frequency and is fixed by the intersection of eq.~\eqref{Eq: par} and the measured DOS; \emph{ii)} 
the parameters $\nu_D$ and $A_{ex}$ of the equation are determined from a fit to the specific heat data, calculated including the TLS contribution. 

\begin{table*}[ht]
\caption{Best fitting parameters and comparison with Brillouin spectroscopy data. The first three columns of the table report the values of the best fitting parameters: $\nu_D$, $A_{ex}$ and $C_{TLS}$ determined by fitting eq.~\eqref{E:Cp} to the specific heat data of the two glasses. The second three columns are the same parameters expressed as: Debye level in the specific heat, $C_D^{DOS}$, coefficient of the $T^5$ term in the specific heat, $C_{ex}$, and density of TLS per unit frequency, $n_{TLS}$. The last two columns report the Debye frequency and the Debye level determined from the elastic moduli measured with Brillouin spectroscopy~\cite{Moratalla2023}, as detailed in the appendix. The parameters $\nu_D$ and $A_{ex}$ for the crystal are estimated by fitting eq.~\eqref{E:CplowT} to the crystal specific heat at low temperature, considering that in the crystal there are no TLS.}
\label{tab:table}
\begin{ruledtabular}
\begin{tabular}{c|ccc|ccc|cc}
& \begin{tabular}[c]{@{}c@{}}$\nu_{D}$\\ {[}THz{]}\end{tabular} & \begin{tabular}[c]{@{}c@{}}$A_{ex}$\\ {[}THz$^{-5}${]}\end{tabular}   & \begin{tabular}[c]{@{}c@{}}$C_{{TLS}}$\\ {[}$\mu$J/gK$^{2}${]}\end{tabular}  & \begin{tabular}[c]{@{}c@{}}$C_{D}^{DOS}$\\ {[}$\mu$J/gK$^{4}${]}\end{tabular}  & \begin{tabular}[c]{@{}c@{}}$C_{ex}$\\ {[}$\mu$J/gK$^{6}${]}\end{tabular} & \begin{tabular}[c]{@{}c@{}}$n_{{TLS}}$\\ {[}THz$^{-1}$atoms$^{-1}${]}$\times$10$^{-5}$\end{tabular} & \begin{tabular}[c]{@{}c@{}}$\nu_{D}^{BLS}$\\ {[}THz{]}\end{tabular} & \begin{tabular}[c]{@{}c@{}}$C_{D}^{BLS}$\\ {[}$\mu$J/gK$^{4}${]}\end{tabular}  \\ \hline
OG  & 4.27 $\pm$ 0.06  &  0.65 $\pm$ 0.03 & 8.1 $\pm$ 0.6  & 32 $\pm$ 1 & 6.5 $\pm$ 0.3 & 20 $\pm$ 1  & 4.21 $\pm$ 0.07 & 33 $\pm$ 2 \\
USG & 4.29 $\pm$ 0.08  &  0.66 $\pm$ 0.06 & 1.5 $\pm$ 0.6  & 31 $\pm$ 2 & 6.6 $\pm$ 0.6 &  4 $\pm$ 2  & 4.64 $\pm$ 0.07 & 24 $\pm$ 1  \\ 
Crystal & 4.421 $\pm$ 0.004 &  0.111 $\pm$ 0.003   &   & 28.35 $\pm$ 0.08 & 1.11 $\pm$ 0.03 & & &    
\end{tabular}
\end{ruledtabular}
\end{table*}
Specifically, eq.~\eqref{E:Cp} with our model for the harmonic DOS is used as a fitting function to the measured specific heat, where we use the DOS at 150 K for the USG and the room temperature one for the OG. 
The best values of the Debye frequency, $\nu_D$, of the $A_{ex}$ coefficient and of the TLS contribution, $C_{TLS}$, estimated with a routine that minimizes the $\chi^2$ between calculated and experimental specific heat, are reported in Table~\ref{tab:table}. The very good agreement between the computed, red lines in Figure~\ref{Figs: Cp}, and measured specific heat evidences the quality of the measured DOSs, which dominate the specific heat at and above the BP maximum.
The low frequency part of the reduced DOSs determined in this way is plotted in Figure~\ref{Figs: dos} as a black line. We are here making the simplified assumption that eq.~\eqref{Eq: par} is valid up to the BP frequency, because a direct measurement of the DOS at the temperatures relevant for the specific heat calculation is unfeasible with the presently available X-ray fluxes. 

Our estimate of $\nu_D$ for the ordinary glass is in agreement within uncertainty with the value determined from the elastic moduli ($\nu_D^{BLS}$ in Table~\ref{tab:table}). The nice agreement between these two independent estimates of the low temperature Debye level supports the reliability of our approach and the use of eq.~\eqref{Eq: par}. 
On the contrary, the Debye frequency that we estimate for the ultra-stable glass is significantly lower than the value determined from the moduli (see Table~\ref{tab:table}). This discrepancy is due to the anisotropic arrangement of TPD molecules during deposition at 0.85 $T_g$, where molecules mostly orient parallel to the substrate~\cite{Moratalla2023,aniso,Javier_aniso}. The sound velocities were measured by Brillouin spectroscopy for waves propagating in a direction parallel to the substrate, giving values higher than the average and an apparently lower Debye level, even lower than that of the crystal (values in Table~\ref{tab:table}). Our analysis also confirms the marked reduction of $C_{TLS}/T^3$ (violet dotted lines in Figure~\ref{Figs: Cp}) as the glass stability is increased from the OG to the USG, although a small but non-negligible density of TLS is still present in the ultra-stable glass.

An unexpected result is, instead, the observation of a similar departure from the Debye level in both the OG and the USG. The parameters $A_{ex}$ of both glasses are equal within uncertainty, implying that the glass stability and the number of two-level states have no measurable effect on the harmonic vibrations at the lowest frequencies, those lying well below the BP. 

At low temperatures, below the BP maximum, the specific heat can be written as~\cite{Ramos2002}:
\begin{equation}\label{E:CplowT}
    C_p \sim C_{TLS} T + C_D T^3 + C_{ex} T^5.
\end{equation}
The parameters $C_D$ and $C_{ex}$ are, respectively, proportional to $1/\nu_D^3$ and $A_{ex}$, with the coefficients reported at the end of appendix~\ref{App:Cp}. Their value is included in Table~\ref{tab:table} for clarity. Equivalent parameters for the crystal can be estimated by a direct fit of the previous equation to the low temperature part of $C_p/T^3$, with $C_{TLS} = 0$. They are reported in Table~\ref{tab:table} and the fit shown in Figure~\ref{Figs: Cp}a) as a green line. The $A_{ex}$ parameter derived for the crystal in this way is almost an order of magnitude smaller than that of the two glasses, indicating that $A_{ex}$ is strongly affected by disorder, although apparently insensitive to the glass stability.\\

\section{Discussion}
\begin{figure}[ht]
    \centering
    \includegraphics[width=0.95\linewidth]{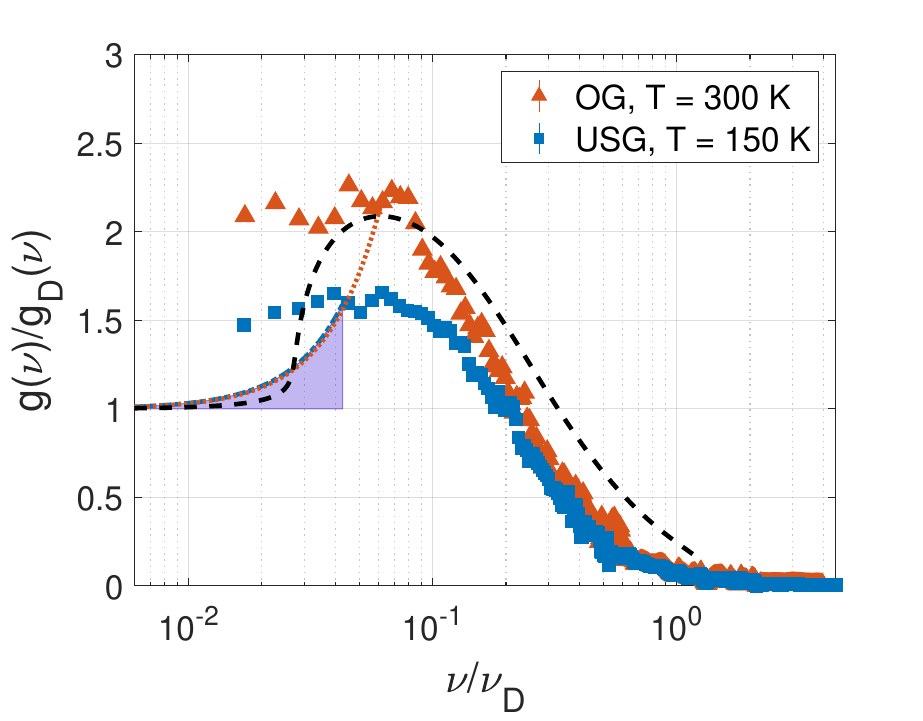}
    \caption{The reduced density of states in Debye units. As indicated in the legend, the red triangles represent the OG measured at 300 K, while the blue squares the USG measured at 150 K. The frequency is reported in units of the Debye frequency, $\nu_D$, and $g(\nu)$ is divided by the Debye DOS, $g_D(\nu) = A_D \nu^2$. The low frequency estimate of the harmonic DOS based on eq.~\eqref{Eq: par} is shown as a dotted red line for the OG and a dash-dotted blue line for the USG. The shaded violet region indicates the contribution of the excess modes, $A_{ex} \nu^4$ in eq.~\eqref{Eq: par}, to the reduced harmonic DOS, highlighting the fact that this contribution is insensitive to the glass stability. The dashed black line is the HET model with parameters optimized to describe the measured DOS, as detailed in appendix~\ref{App:HET}.}
    \label{Figs: dos DL}
\end{figure}

Having determined the proper values of the Debye levels, we can now plot the DOS in Debye units, as shown in Figure~\ref{Figs: dos DL}. The figure allows one to appreciate the strong effect of the glass stability on the BP intensity. The BP decreases by almost a factor of two in passing from the OG to the USG, confirming similar observations in numerical studies of vapor deposited glasses~\cite{Singh2013} and in experiments on stable polymer spheres~\cite{Monnier2021}. However, the lower frequency portion of the estimated $g(\nu,0)$, the shadowed region in the figure, does not change in an appreciable way with glass stability. It is worth noting that our estimate of the low temperature DOS is quite robust because it is based on the low temperature specific heat data, as extensively described above.

We now compare the measured data with the predictions of the soft potential model~\cite{Karpov1983,Buchenau1991,Parshin2007} and of the heterogeneous elasticity theory~\cite{HET_ruocco,articlemodel,SCHIRMACHER2015133,PhysRevB.104.134106}, two well established and competing models for the terahertz vibrations of glasses. It is worth recalling that these models have a phenomenological nature, since they are not grounded in a microscopic description of the chosen glass. The advantage of these approaches is that their predictions can be compared to experimental data, at the expense of one or more adaptable parameters.

In a second part of this discussion section, we focus on the effect of sound attenuation on the departure of the harmonic DOS from the Debye level and on the origin of the anharmonicity of the DOS at low frequencies.

\subsection{The soft potential model}
The soft potential model~\cite{Karpov1983,Buchenau1991,Parshin2007,Ramosbook,Buchenaubook} is an extension of the standard tunneling model to higher temperatures and frequencies. It assumes the presence in the glass of a distribution of soft an-harmonic potentials, that give rise to two level states and to quasi localized modes. The QLM dominate the DOS at and below the BP frequency, with a characteristic density that grows as the fourth power of frequency up to the BP maximum, as in the previously cited numerical simulations~\cite{Lerner,Angelani,Mizuno,Shcheblanov2020,Bouchbinder2020,Bouchbinder2023}. The model has three free parameters: the density of soft modes, $P_s$, a frequency, $W/h$, that separates the TLS states at low frequency from the harmonic vibrations at higher frequency and a coefficient, $\Lambda$, that describes the coupling of the QLM with the sound waves. This third coefficient is necessary only for calculations of sound attenuation and of thermal conduction, but it is not relevant for the DOS and the specific heat. 

If we attribute the departure from the Debye level in the harmonic DOS entirely to the presence of QLM, as normally done in this model, we obtain the parameters reported in table~\ref{tab:SPM} of appendix~\ref{App:SPM}, where details of the calculation are reported.
The soft potential model can thus adapt to the present situation, where the glass stability strongly influences the density of TLS but does not affect the density of QLM. 
In addition, the model gives a prediction for the relaxational part of the DOS, $g_{rel}(\nu,T)$ in eq.~\eqref{eq:relvib}, once the parameters $P_s$ and $W$ are known. The prediction overestimates by approximately a factor of two the observed low frequency part of the measured $g(\nu,T)$, as shown in Figure~\ref{Figs/DOSrel} in appendix~\ref{App:SPM}. The failure of the model to quantitatively describe the temperature dependence of the DOS can be explained in two different ways: \emph{i)} thermally activated relaxation processes could saturate as the frequency is increased to the terahertz range. \emph{ii)} the weight of the QLM could be overestimated. We believe this second explanation is more appropriate and consistent with our measurements, as we discuss below in the section on the anharmonic DOS. 

As concerns the vibrational modes at higher frequencies, the soft potential model, in its original formulation, describes the departure from the Debye level but not the BP maximum. Extensions of the model to deal with the excess of modes at the BP have been introduced later on~\cite{PhysRevLett.70.182,Parshin2007}. However, the extended model does not include the contribution of the non localized vibrational modes, sound waves and high frequency vibrations, so that it is impossible to compare it with the measured data without resorting to additional assumptions that appear unjustified~\cite{Buchenau2007}. In this sense the soft potential model is inadequate to explain the observation of a strong sensitivity of the BP to the glass stability.

\subsection{The heterogeneous elasticity theory}
The heterogeneous elasticity theory (HET)~\cite{HET_ruocco,articlemodel,SCHIRMACHER2015133,PhysRevB.104.134106} predicts the appearance of the BP as a result of the microscopic spatial fluctuations of the local shear modulus. It describes the terahertz vibrations of glasses in a harmonic approach where the vibrations are unrelated to the instabilities associated with two level states. The DOS in the HET model is computed by means of a ``generalized" Debye law~\cite{HET_ruocco} where the first departure from the Debye level, giving rise to the coefficient $A_{ex}$ in eq.~\eqref{Eq: par}, is associated to the dispersion and damping of the sound waves (see eq.~\eqref{Eq: model g(w)} in appendix~\ref{App:HET}).
An-harmonic processes can modify the sound dispersion and attenuation, thus giving rise to a temperature dependence of the DOS~\cite{Baldi2014}, while the eventual effect of secondary relaxation processes is not considered.

We compute here the HET prediction in the limit of small disorder, where a single free parameter, $\gamma$, is used to describe the variance of the local shear modulus fluctuations (see appendix~\ref{App:HET} for details on the model calculation and the used parameters). The calculation (dashed line in Figure~\ref{Figs: dos DL}) provides a qualitatively good description of the DOS of the ordinary glass, despite some quantitative disagreement on the high-frequency side. This discrepancy is reflected in an overestimation of the specific heat above the BP temperature, as we discussed recently in ref.~\cite{festi}. 
The failure of the model at high frequencies is most probably due to the fact that the generalized Debye law forces all the frequencies to lie below the Debye frequency, $\nu_D$, which is an order of magnitude smaller than the maximum vibrational frequency in TPD (see Figure~\ref{Figs/Analysis: ext} for the full spectrum).
The main limit of the HET model is, however, its inability to describe the low BP intensity we observe in the USG glass. HET predicts a BP intensity which is twice the value of the Debye level for high values of the disorder parameter, as needed to describe a BP located at low frequencies as in these glasses~\cite{HET_ruocco}, while in the USG the excess is of only 50\%. 

\subsection{Contribution of sound attenuation to the DOS}
The generalized Debye law exploited in the HET, and in similar theories based on a random matrix approach~\cite{Grigera2003,Vogel2023}, can be used to estimate the contribution of the sound attenuation to the DOS, independently of the approximations introduced in the various models. In the small frequency limit, the relation simplifies to~\cite{HET_ruocco,Flenner2025}:
\begin{equation}\label{eq:DOSsound}
    g(\nu) \sim \frac{3 \nu^2}{\nu_D^3} + \frac{2 v_D^2}{\pi \nu_D^2} \left[ \frac{\Gamma_L}{v_L^2} + 2\frac{\Gamma_T}{v_T^2} \right],
\end{equation}
where $v_{L,T}$ and $\Gamma_{L,T}$ are the sound velocities and sound attenuation (full width at half maximum) of longitudinal ($L$) and transverse ($T$) waves, and $v_D$ is the Debye sound velocity.
The sound attenuation in glasses follows the Rayleigh scattering law, $\Gamma_{L,T} \sim A_R \nu^4$, at sufficiently low temperatures and frequencies~\cite{Baldi2010,Baldi2014}, so that eq.~\eqref{eq:DOSsound} reduces to eq.~\eqref{Eq: par} with $A_{ex} \propto A_R$. 
If we attribute the departure from the Debye level entirely to the damping of sound waves, thus assuming the absence of QLM, and we consider a comparable damping of longitudinal and transverse modes, we can compute the frequency, $\nu_{IR}$, of the Ioffe-Regel limit, where the amplitude mean free path of the sound waves is comparable to a wavelength~\cite{Baldibook}. We find $\nu_{IR} \sim 0.36$ THz for both the OG and the USG, a value very close to the BP position, as observed in many glasses~\cite{Ruffle2006}.

\begin{figure}[ht]
    \centering
    \includegraphics[width=0.95\linewidth]{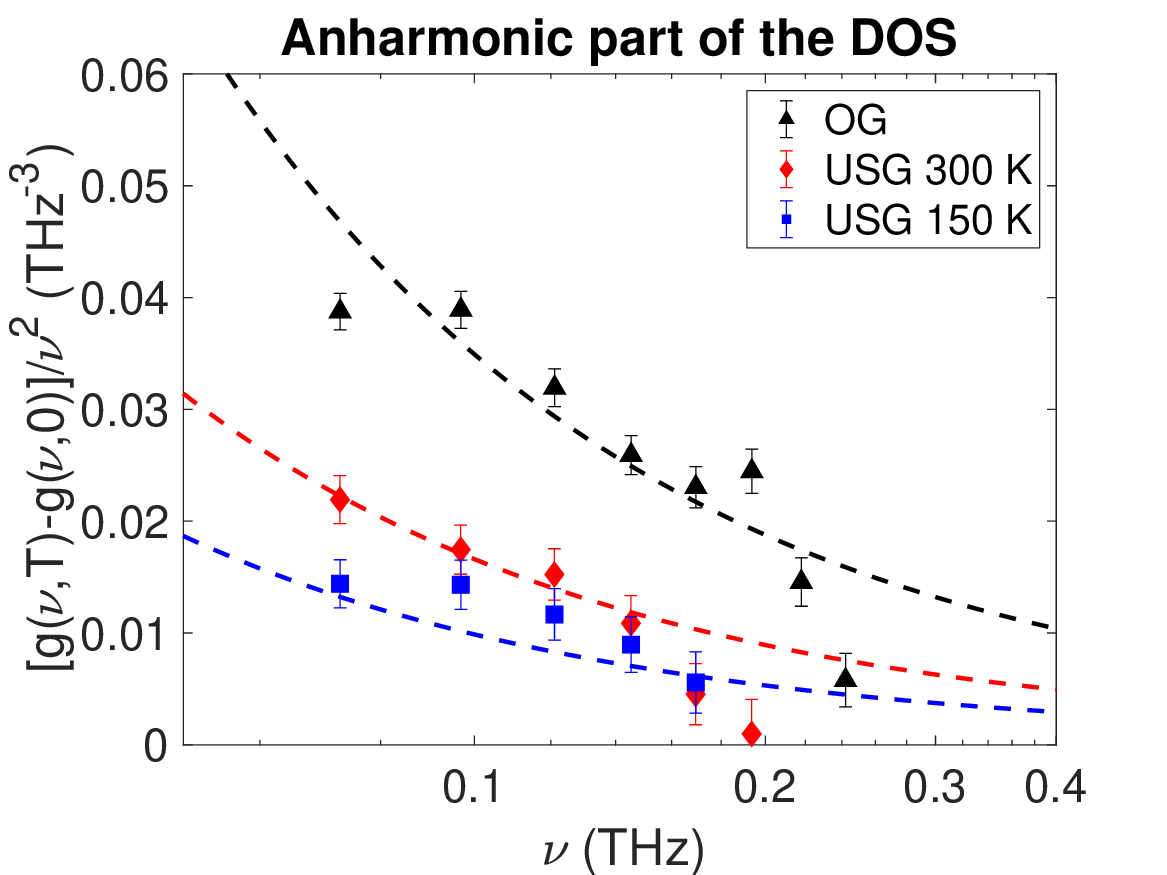}
    \caption{Anharmonic part of the reduced density of states, calculated as difference of the measured reduced DOS and of its harmonic contribution, modeled by eq.~\eqref{Eq: par}. Only positive values of this quantity are plotted. The lines are fits of eq.~\eqref{eq:SPMrel} with only $W$ as free parameter, since the ratio $P_s/W$ is fixed by the knowledge of $C_{TLS}$. The $W$ parameter for the USG is fitted to the room temperature data.}
    \label{Figs: dos Anh}
\end{figure}
\subsection{Anharmonic part of the density of states}
The soft potential model has been used with some success to describe the anharmonic part of the density of states of various glasses~\cite{Buchenau2007}, with a relaxational component of the DOS given by the following expression:
\begin{equation}\label{eq:SPMrel}
    g_{rel}(\nu,T) = \frac{1}{3} \frac{P_s h}{W^2} h \nu \left( \frac{k_B T}{W} \right)^{3/4} ln^{-1/4} \left( \frac{1}{2 \pi \nu \tau_0} \right) ,
\end{equation}
where $\tau_0 \sim 0.1$ ps. This temperature dependent contribution implies that the reduced DOS, $g(\nu)/\nu^2$, diverges as $\sim \nu^{-1}$ at low frequencies at finite temperature, since the logarithmic term is approximately constant in the relevant frequency range. The ratio $P_s/W$ is fixed by the soft potential expression for the TLS contribution to the specific heat, $C_{TLS}$, given by equation~\ref{eq:SPMTLS} in appendix~\ref{App:SPM}. 

\begin{table}[ht]
\centering
\caption{Soft potential model parameters, $P_s$ and $W$, for the USG and the OG, with $W$ optimized from a fit of eq.~\eqref{eq:SPMrel} to the difference between the reduced DOS and its harmonic component. The third column is the value of the $A_{ex}$ parameter in eq.~\eqref{Eq: par} predicted by soft potential model. This value, $A_{ex}^{QLM}$, is the coefficient of the $\sim \nu^4$ dependence of the density of QLMs from eq.~\eqref{eq:SPMdos}.}
\label{tab:SPManh}
\begin{tabular}{c|c|c|c}
  \hline
  \hline
 &
  \begin{tabular}[c]{@{}c@{}}$P_s$\\ {[}state/atoms{]}$\times$10$^{-6}$\end{tabular} &
  \begin{tabular}[c]{@{}c@{}}$W/h$\\ {[}GHz{]}\end{tabular} &
  \begin{tabular}[c]{@{}c@{}}$A_{ex}^{QLM}$\\ {[}THz$^{-5}${]}\end{tabular} \\
  \hline
OG & 2.3 $\pm$ 0.1   & 70 $\pm$ 1  &  0.057 $\pm$ 0.004  \\ \hline
USG  & 0.3 $\pm$ 0.1   & 43 $\pm$ 6  &  0.08 $\pm$ 0.06 \\ 
  \hline
  \hline
\end{tabular}
\end{table}
We can estimate the parameter $W$ that better describes the anharmonic part of the DOS, if we assume that the difference between the measured DOS and the harmonic DOS is entirely attributable to $g_{rel}(\nu,T)$. This is a quite strong assumption, since the classical hopping in double well potentials gives rise simultaneously to anharmonic damping of sound and to the relaxational component of the DOS. Consequently, this mechanism, together with possible other anharmonic processes, may give a sizable contribution to the DOS via the sound attenuation and eq.~\eqref{eq:DOSsound}. However, the absence of data for the sound attenuation of TPD prevents us to include these possible additional effects in our analysis. The best fit of eq.~\eqref{eq:SPMrel} to the quantity $\left[g(\nu,T)-g(\nu,0) \right]/\nu^2$ is reported in Figure~\ref{Figs: dos Anh} and the corresponding soft potential model parameters are listed in table~\ref{tab:SPManh}. The agreement between the estimate of the anharmonic part of the DOS and the data is quite good, failing only on the high frequency side of the figure, where both eq.~\eqref{Eq: par} and~\eqref{eq:SPMrel} are unreliable. In particular, the difference between the two temperatures measured for the USG glass is well captured by the model.

Having determined the parameters $P_s$ and $W$ we can use the soft potential model to predict the density of quasi localized modes, from eq.~\eqref{eq:SPMdos} of appendix~\ref{App:SPM}. The resulting estimate of the contribution of QLM, $A_{ex}^{QLM}$, to the parameter $A_{ex}$ is reported in the last column of table~\ref{tab:SPManh}. The QLM contribution appears negligible with respect to the other mechanisms that can give rise to a departure from the Debye law, since the values of $A_{ex}^{QLM}$ are an order of magnitude smaller than the measured $A_{ex}$ for both samples (see table~\ref{tab:table}). This suggests that the major component contributing to the departure of the harmonic DOS from the Debye law is given by the attenuation of sound, as explained in the previous section. The bending of the dispersion curve, responsible for this deviation in crystals, gives a small contribution in the studied samples, since the $A_{ex}$ for the crystal is small, as discussed in section~\ref{S:harm}. It is worth noting that the value of $A_{ex}^{QLM}$ for the USG is affected by a significant uncertainty, mainly due to the uncertainty on $C_{TLS}$, which is close to zero in this glass. Consequently, the experimental precision doesn't allow us to check whether QLM have a lower density in the more stable glass, as predicted by the numerical works.

\section{Conclusions}
We have reported on the measurement of the DOS of an ultra-stable glass and of the corresponding ordinary sample, using the nuclear resonant analysis of inelastic X-ray scattering with a spectrographic approach. The innovative method allows us to probe the DOS of the amorphous thin films in absolute units from a minimum frequency of $\sim 0.07$ THz, giving access to the frequency range where quasi localized harmonic vibrations (QLM) are predicted by the soft potential model and observed in numerical simulations.
The low frequency part of the measured DOS is temperature dependent and lies above the Debye levels determined from the elastic moduli, indicating the presence of an-harmonic or relaxational effects, superimposed to the harmonic DOS inferred from specific heat data. We show that the measured DOS is in quantitative agreement with the specific heat and discloses a significant elastic anisotropy in the ultra-stable glass, related to the anisotropic packing of molecules in this sample. 

The comparison between the ordinary and the ultrastable glass allows us to reveal that the departure of the harmonic, low temperature, DOS from the Debye law is almost insensitive to the glass stability.
On the contrary, the BP intensity reduces markedly as the glass stability is increased.
We attribute the departure of the reduced harmonic DOS from the Debye level to the damping of sound waves, due to Rayleigh scattering. An estimate of the Ioffe-Regel limit based on this assumption indicates that it is located very close to the BP frequency, in-line with previous observations on other glasses~\cite{Ruffle2006}. 
The temperature dependence of the DOS at low frequencies can be quantitatively described within the soft potential model (SPM), which postulates that QLM emerge as high frequency counterparts of the low frequency excitations within two level states. The transition between the two families of excitations takes place at a frequency that is a function of glass stability, possibly reflecting the different shape of the energy landscape in the different minima. 
A quantitative description of the an-harmonic part of the DOS implies that the density of QLM is an order of magnitude smaller than the value required to explain the measured $\sim \nu^4$ increase of the harmonic DOS, supporting the relevance of sound attenuation to this increase.
The strong sensitivity of the BP to the glass stability, in line with numerical simulation results~\cite{Singh2013} is, instead, not captured by the SPM, because of its known limitations in describing the vibrational modes at the BP.

The relevance of the contribution of sound attenuation to the deviation of the DOS from the Debye law can also explain the apparent disagreement between our observations and the numerical simulations~\cite{Wang_2019,flenner202504,Flenner2025}, concerning the sensitivity of the low frequency vibrations to the glass stability.
The samples under study have the low frequency DOS dominated by the contribution of propagating vibrational modes and of their damping. The contribution of QLM is subdominant with respect to this mechanism, so that the variation of the QLM density with stability is probably hidden by the presence of phonon-like excitations.
A direct measurement of sound damping as a function of glass stability, in the frequency range where it is dominated by Rayleigh scattering, would allow us to definitively clarify the relevance of sound attenuation with respect to the presence of QLM in different classes of amorphous solids. Methods to probe Rayleigh scattering in thin films in the relevant frequency range have been recently developed, and the feasibility of the measurement has been demonstrated for vitreous silica~\cite{Fainozzi2024,Wang2025}. The combined measurement of the low frequency tail of the DOS and of the Rayleigh scattering of sound holds the promise to allow us to clarify the nature of the low frequency vibrations in glasses.

\section{Acknowledgements}
We thank Mr. J.-P. Celse for providing technical assistance during the beamtime at ID18 at the European Synchrotron Radiation Facility (ESRF). We acknowledge ESRF for provision of synchrotron radiation facilities under proposals number HC-4933 and HC-5322 at beamline ID18 (now ID14). F. C. is a charg\'{e} de recherche of the Fund for Scientific Research (F.R.S.–FNRS). M. M. and M. A. R. acknowledge financial support from the Spanish Ministry of Science, Innovation and Universities (MCIN/AEI/10.13039/501100011033) through the grant PID2021-127498NB-I00, and within the “María de Maeztu” Program for Units of Excellence in R\&D (CEX2023-001316-M). J. R.-V., C. R.-T. and M. R.-L. acknowledge grants PID2020-117409RB-I00, PID2023-147645NB-I00 from MICIU and 2021SGR-00644 funded by AGAUR. M. R.-L. was in receipt of a grant from AGAUR. The ICN2 is funded by the CERCA programme/Generalitat de Catalunya and supported by the Severo Ochoa Centres of Excellence programme, grant no. CEX2021-001214-S, funded by grant no. MCIN/AEI/10.13039.501100011033.

\section{Author contributions}
G.B. and J.R.V. conceived and designed the project. E.A., G.B., D. B., A.I.C., F.C., I.F and C.R.T. performed the experiments. Sample fabrication was carried out by M.R.L. and C.R.T.. I.F. analyzed the data with support from A.I.C. and F.C. G.B. and I.F. performed the model calculations, with the help of M.A.R.. G.B. and I.F. wrote the manuscript with contributions from all the authors. All the authors contributed to the discussion of the results, reviewed the data and provided feedback on the manuscript.


\appendix

\section{Methods}
\label{Methods}
\subsection{Samples}
\label{Methods: samples}
The samples studied consist of ultra-stable and conventional TPD glass films (N,N'-Bis(3-methylphenyl)-N,N'-diphenylbenzidine), prepared at the Universitat Autònoma de Barcelona. TPD is an organic semiconductor with chemical formula C$_{38}$H$_{32}$N$_2$. The films are grown to a thickness of approximately 60 $\mu$m using physical vapor deposition (PVD) onto a $1\times1$ cm$^2$ Si substrate, inside a high-vacuum chamber with a base pressure of $1\times10^{-8}$ Torr. The Si substrate is held at 283 K (0.85$T_g$) during the deposition of the ultra-stable glass and at the glass transition temperature $T_g = 333$ K for the preparation of the ordinary glass. Differential scanning calorimetry (DSC) is used to determine the onset temperatures of the ultra-stable glass, $T_{on} = 363.3$ K. The DSC analysis is performed by heating the as-deposited sample at a rate of +10 K/min, followed by cooling at -10 K/min, as detailed in \cite{Rodriguez-Tinoco2022}. The low temperature specific heat of equivalent samples has been recently investigated by some of us, showing the almost complete disappearance of the TLS contribution in the USG~\cite{Moratalla2023}.

\subsection{NRAIXS measurements}
The density of vibrational states (DOS) of ultra-stable and conventional TPD glasses is measured at room temperature at the Nuclear Resonance beamline~\cite{Rüffer1996} ID18 (now ID14) of the European Synchrotron Radiation Facility (ESRF). The experiments exploit a newly developed spectrograph for inelastic X-ray scattering~\cite{PhysRevLett.123.097402} combined with nuclear resonance analysis (NRAIXS)~\cite{PhysRevLett.76.4258}, achieving an energy resolution (FWHM) of $\sim 130 \mu$eV $\sim 32$ GHz. The measurements are performed by tuning the input X-ray energy within a narrow range (about 15 meV $\sim 3.6$ THz) around the $^{57}$Fe nuclear transition energy (14413 eV), using a spectrograph composed of three asymmetrically cut silicon crystals. We refer to the spectra obtained with this setup as the \textit{high-resolution spectra} (HR spectra).

\begin{figure}[ht]
    \centering
    \includegraphics[width=0.7\linewidth]{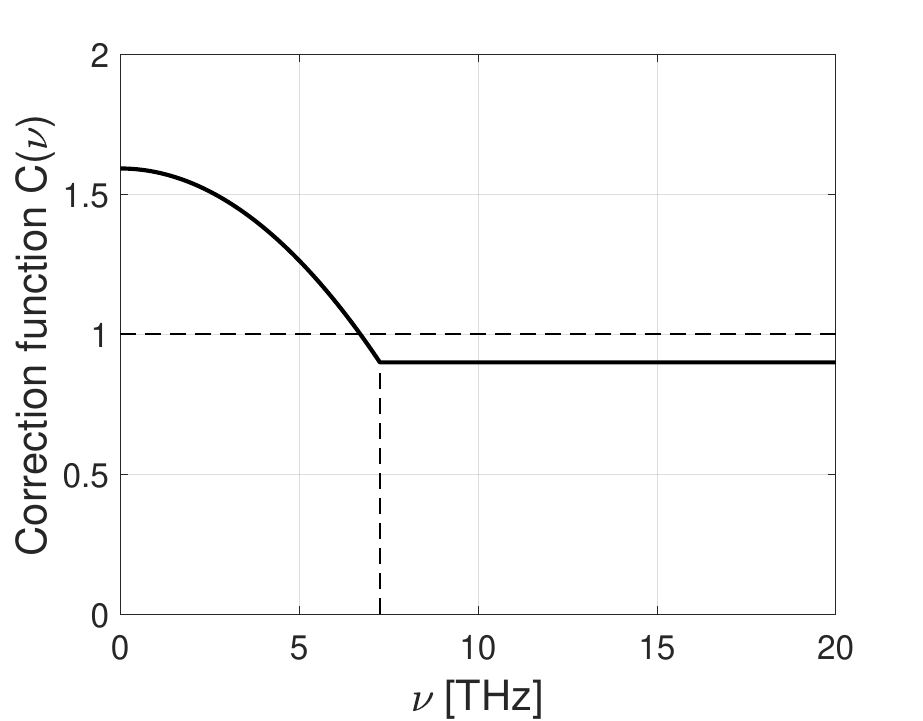}
    \caption{The shape of the correction function used for the TPD molecule.}
    \label{Figs/Analysis: CF}
\end{figure}
To prevent any contribution from the substrate, the X-ray beam crosses the sample in transmission geometry. The focal spot at the sample position is $\sim 30\times$600 $\mu$m$^2$ (V$\times$H), with the vertical size being smaller than the sample thickness. The radiation scattered by the glass films is measured by a detector placed 1 mm above the sample. The transmitted X-ray beam is collected by a similar detector positioned in a forward direction to measure the instrumental response.
The detectors, consisting of a large-area avalanche photodiode covered by a 10 $\mu$m thick $^{57}$Fe foil, act as nuclear resonance filters. Using the Mössbauer effect, the scattered photons are detected at a fixed energy with a resolution of approximately 0.5 $\mu$eV $\sim 0.12$ GHz, negligible compared to the bandwidth of the input beam.
The input X-ray energy is varied in steps of 0.02 meV $\sim 0.005$ THz around the $^{57}$Fe nuclear transition energy, over a range of [-1.2, 2.4] THz. At room temperature, each energy point is measured for 2 seconds, with a total of 65 scans for the ultrastable glass (USG) and 33 scans for the conventional glass. 
The DOS of an ultra-stable TPD glass sample is also measured at 150 K during a second beamtime, to investigate the temperature dependence and low-energy anharmonic contributions (below 0.2 THz). For these measurements, the sample is placed in a helium exchange-gas cryostat. The scanned energy range is [-1.5, 2.4] THz, with a frequency step of 0.005 THz, and each point is measured for 2 seconds. A total of 168 scans are collected.

\begin{figure*}[ht]
    \centering
    \includegraphics[width=1\textwidth]{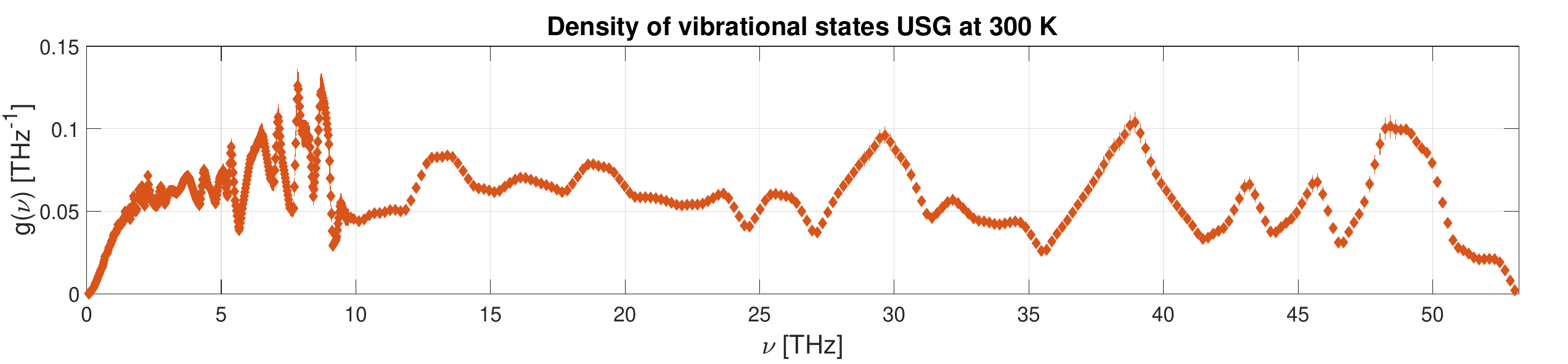}
    \caption{Density of vibrational states $g(\nu)$ measured on the USG at 300 K.}
    \label{Figs/Analysis: ext}
\end{figure*}
Furthermore, to accurately assess the DOS, an extended energy spectrum is measured for the ultra-stable glass at room temperature, spanning a range of [-5, 100] THz. For this, the spectrograph was replaced by a high-resolution monochromator (HRM) with an energy resolution of 1.8 meV $\sim$ 0.45 GHz. Due to time constraints, this extended spectrum is measured only for the ultra-stable glass. It was unnecessary to repeat the measurement at 150 K, as high-energy vibrational spectra follow Bose-Einstein statistics and scale with temperature.
Throughout the experiments, the structure factor of the glasses is periodically checked using $\theta-2\theta$ geometry to monitor the sample condition. The static structure factor remains stable, indicating that the sample is not affected by the X-ray radiation during these measurements. In total, two weeks of beamtime were required to complete both sets of experiments.

\subsection{Data analysis}
\label{Methods: data analysis}
The data analysis procedure is based on the direct method to evaluate the DOS from the energy spectrum of nuclear resonant inelastic scattering \cite{article}. The process begins with the subtraction of the elastic contribution from the raw high-resolution (HR) spectra. The resulting data are then merged with the broader, lower-resolution spectrum.
The multiphonon contribution is removed using Lipkin’s rules and a double Fourier transform. This step requires fixing the mean recoil energy, $E_R$, which is treated as a variable parameter. Initially, $E_R$ is estimated based on the mean atomic mass and the static structure factor ($E_R \simeq 25$ meV $\sim$ 6.1 GHz). Then, it is fine-tuned to ensure that the DOS baseline is consistent with zero at high energies. The resulting recoil energy values are as follows:

\begin{table}[ht]
\centering
\begin{tabular}{c|c|c}
  \hline
  \hline
 &
  \begin{tabular}[c]{@{}c@{}}$E_R$\\ {[}meV{]}\end{tabular}  &
  \begin{tabular}[c]{@{}c@{}}$E_R$\\ {[}THz{]}\end{tabular}\\ \hline
USG - 300 K  & 10.7 & 2.59 \\ 
OG - 300 K  & 10.1 & 2.44  \\ 
USG - 150 K & 18.6 & 4.50  \\
  \hline
  \hline
\end{tabular}
\end{table}
The difference between the recoil energies at 300 K and 150 K is reasonable because, in the low-temperature measurement, the detector is positioned slightly farther away from the sample due to the cryostat. As a result, smaller scattering angles are less represented, leading to a higher recoil energy.

Finally, the resulting DOS, $\tilde{g}(\nu)$, is divided by a correction function $C(\nu)$ to obtain the true DOS, $g(\nu)$. The DOS obtained from the previous procedure is the generalized one since the measured spectra are the sum of the contributions from atoms with very different masses, each weighted by unequal coefficients, which take into account the different atomic cross-sections as well \cite{PhysRevB.80.094303}. Since the correction function for the TPD molecule is unknown, we construct it as follows. In the range of the optical modes (above $\sim$ 30 meV $\sim$ 7.25 GHz), we assume a constant value of $C(\nu) = 0.9$. In the range of the acoustical modes (below $\sim$ 7.25 GHz), we use a parabolic function $C(\nu) = a\nu^2 + b$. The parameter $a$ of the parabola is determined by a scaling coefficient $\eta$ based on the relation \cite{PhysRevB.80.094303}:
\begin{equation}
    \lim_{\nu\to 0} \frac{\tilde{g}(\nu)}{\nu^2} = \eta \lim_{\nu\to 0} \frac{g(\nu)}{\nu^2}
\end{equation}
and 
\begin{equation}
    \eta \simeq \frac{\langle m^2\rangle}{\langle m\rangle^2} \ \ ,
\end{equation}
where $m$ is the atomic mass. In the TPD case $\eta \simeq 1.59$. The parameter $b$ instead is set to ensure the continuity between the parabolic function and the constant value of 0.9 at the transition frequency from acoustic to optical modes $\nu_a = 7.25$ GHz. The resulting correction function is shown in Figure \ref{Figs/Analysis: CF}. The choice of $C(\nu)$'s functional form and the transition frequency $\nu_a$ is guided by previously evaluated $C(\nu)$ functions for other systems (see Fig. S9 of \cite{PhysRevLett.112.025502}). However, the specific form of $C(\nu)$ has little impact on the main results.

We report in Figure \ref{Figs/Analysis: ext} the density of vibrational states $g(\nu)$ of the USG measured at 300 K in the entire frequency range: [0 - 53] THz.

\section{Specific heat calculation}\label{App:Cp}
The specific heat of the OG and USG glasses can be quantitatively determined from the measured DOS, following the method discussed in the main text. Here we provide additional details of the calculation. The calorimetric measurements were conducted at the Universidad Autónoma de Madrid with $^3$He and $^4$He cryostats, employing the thermal relaxation method. The samples were placed on a sapphire substrate within the calorimetric cell. Further experimental details are provided in \cite{Moratalla2023}.

In the harmonic approximation, the specific heat per unit mass, $C_p(T)$, is related to the DOS by the following relation:
\begin{equation}\label{eq:Cp}
    C_p(T) =  \frac{3}{M_{at}} \int d\nu g(\nu) \frac{\partial n(T,\nu)}{\partial T} h\nu \ ,
\end{equation}
where $k_B$ is the Boltzmann constant and the DOS is normalized to unity: $\int_0^{\infty} g(\nu) d\nu =1$. The Bose population factor for phonons is:
\begin{equation}
    n(T,E) = \frac{1}{\text{e}^{\frac{h\nu}{k_BT}} -1} \ .
\end{equation}

\begin{figure*}[ht]
    \centering
\subfloat[\label{Figs/Analysis: rdos usg}]{%
  \includegraphics[width=0.3\textwidth]{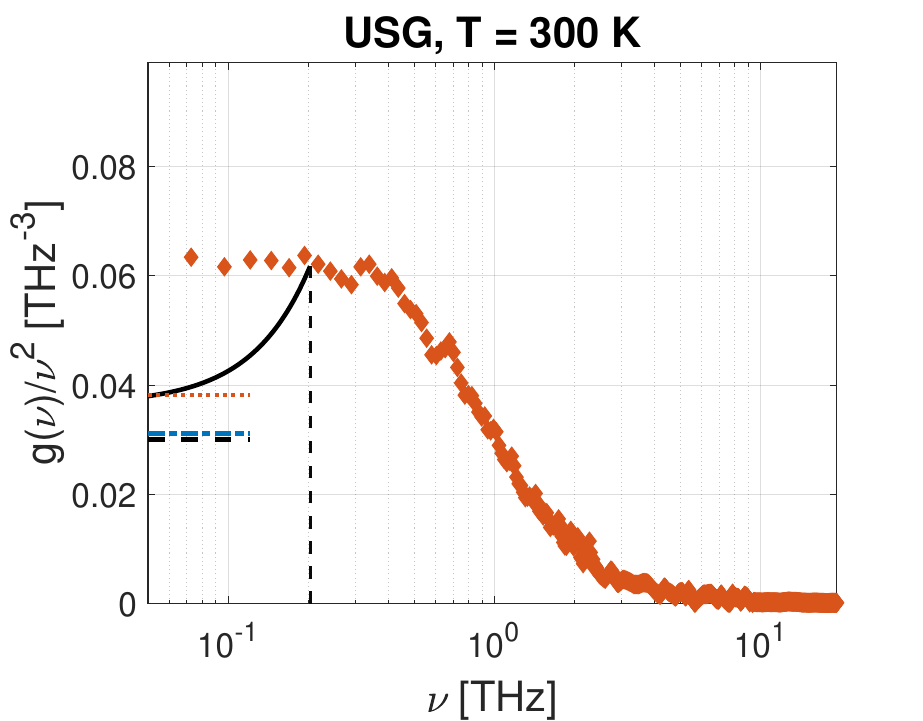}%
}\hspace*{\fill}%
\subfloat[\label{Figs/Analysis: rdos og}]{%
  \includegraphics[width=0.3\textwidth]{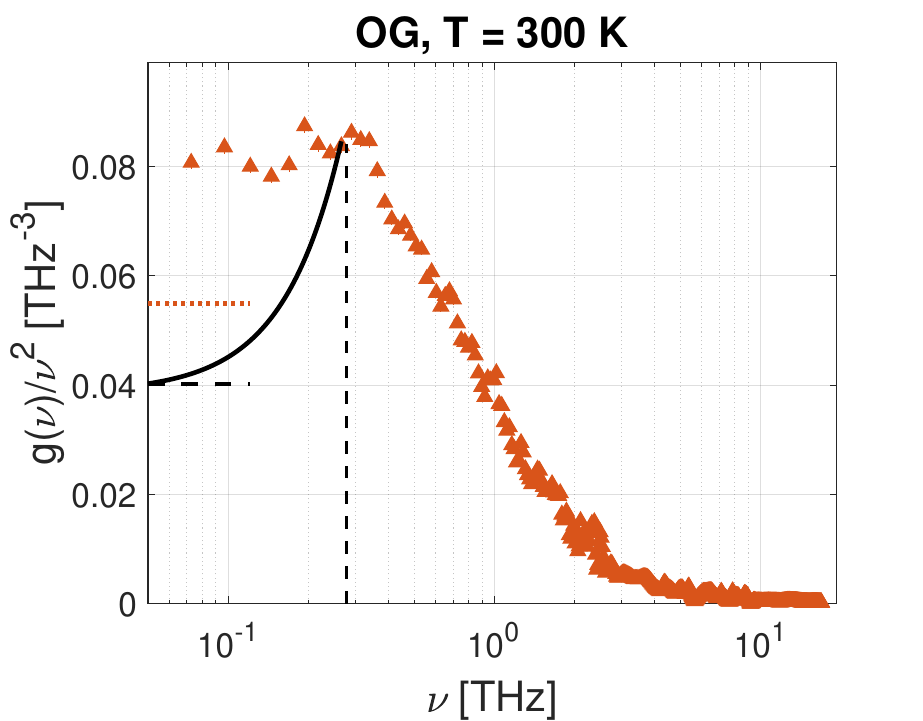}%
}\hspace*{\fill}%
\subfloat[\label{Figs/Analysis: rdos 150}]{%
  \includegraphics[width=0.3\textwidth]{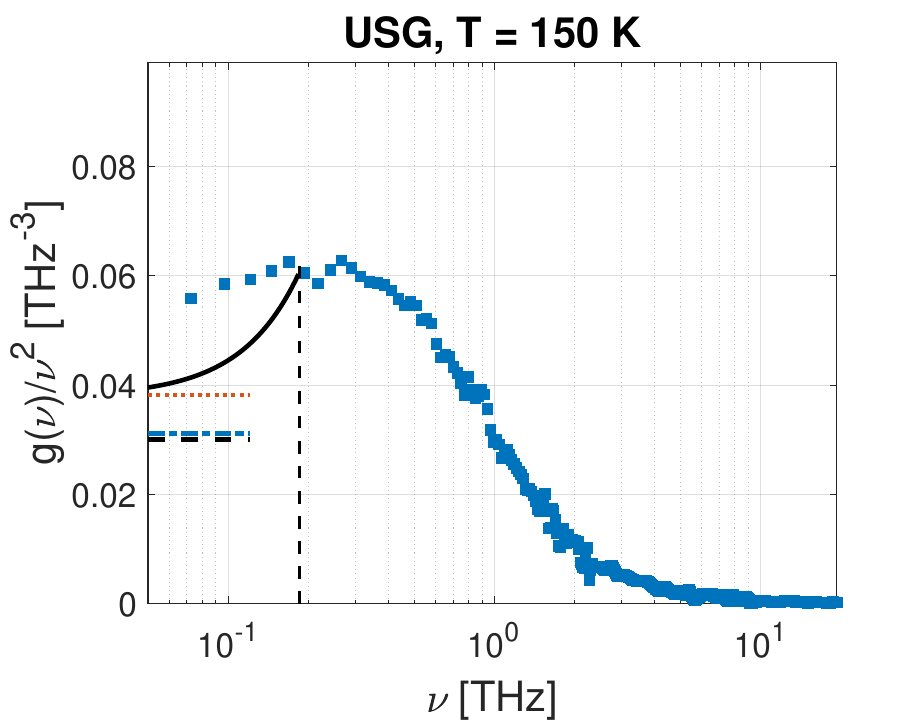}%
}
\caption{The reduced DOS for the three cases. The reduced DOS for the USG measured at 300 K (panel a)), the OG measured at 300 K ( panel b)) and the USG measured at 150 K (panel c)). The points of the DOS below $E_0$ (the black dashed vertical lines) are excluded from the $C_p$ evaluation and approximated with the parabolic curve (black solid lines). The Debye levels for the two samples calculated from the elastic moduli measured with Brillouin spectroscopy at 300 K, 150 K and $T \rightarrow 0$ K are depicted as horizontal lines respectively in red, blue and black.} \label{Figs/Analysis: rdos}
\end{figure*}
To improve the specific heat calculation, we exclude the lower frequency points from the reduced DOS and we approximate them with a parabolic curve (see main article). The low frequency parabolic extrapolation of the reduced DOS, $g_{LF}(\nu)/\nu^2$, goes from the Debye level, $3/\nu_D^3$, where $\nu_D$ is the Debye frequency, to the first intersecting point with the experimental DOS (Figures (\ref{Figs/Analysis: rdos usg}), (\ref{Figs/Analysis: rdos og}), (\ref{Figs/Analysis: rdos 150})):
\begin{equation}\label{eq:parabola}
    \frac{g_{LF}(\nu)}{\nu^2} = \frac{3}{\nu_D^3} + A_{ex}\nu^2 .
\end{equation}
We need to impose the continuity between the parabolic function and the measured DOS at the intersecting frequency value $\nu_0$, defined as:
\begin{equation}\label{eq:constraint}
    \frac{g(\nu_0)}{\nu_0^2} = \frac{3}{\nu_D^3} + A_{ex}\nu_0^2 \ .
\end{equation}
The frequency $\nu_0$ represents the value below which the points of the measured DOS are substituted by the parabolic extrapolation in the specific heat calculation.

\begin{figure*}[ht]
    \centering
\subfloat[\label{Figs/Analysis: Cp_fit3 usg}]{%
  \includegraphics[width=0.3\textwidth]{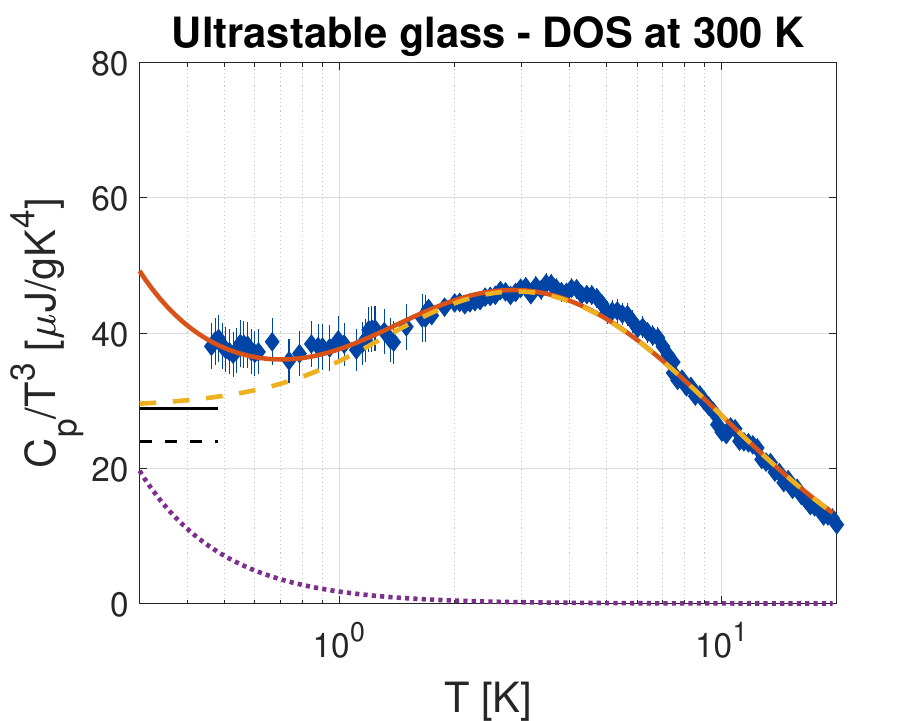}%
}\hspace*{\fill}%
\subfloat[\label{Figs/Analysis: Cp_fit3 og}]{%
  \includegraphics[width=0.3\textwidth]{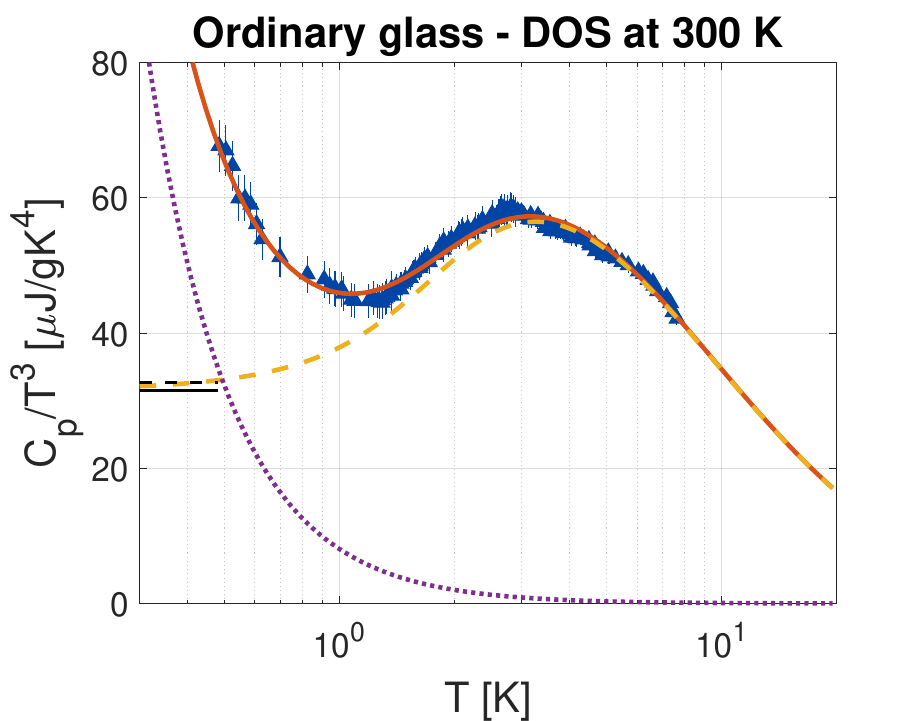}%
}\hspace*{\fill}%
\subfloat[\label{Figs/Analysis: Cp_fit3 150}]{%
  \includegraphics[width=0.3\textwidth]{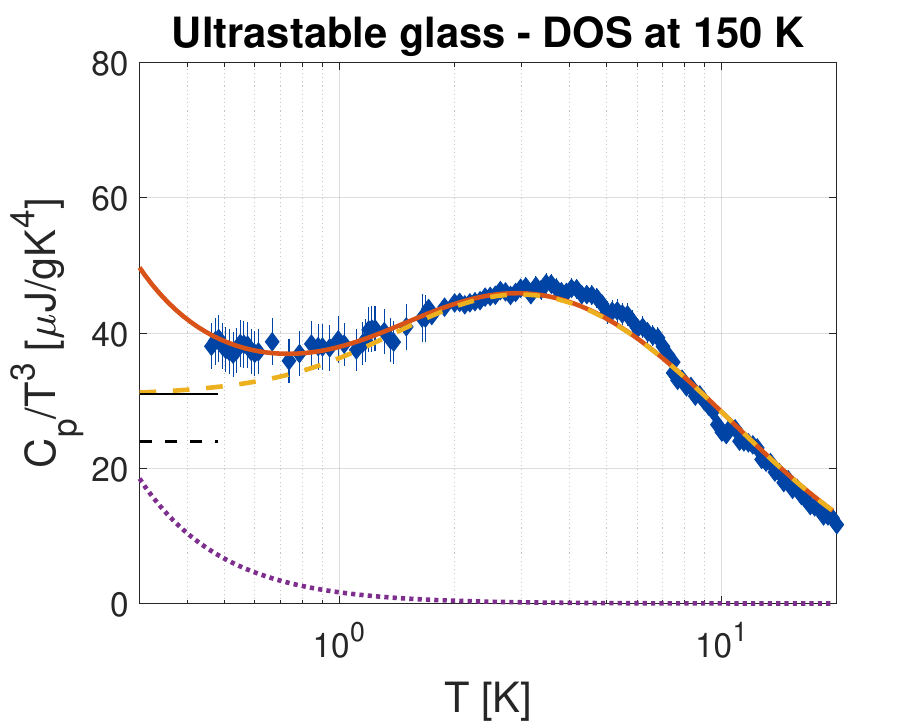}%
}
\caption{The specific heat over $T^3$ and the fit for the three cases. The experimental $C_p/T^3$ compared to the $C_p$ evaluated from the DOSs of the USG measured at 300 K (panel a)), the OG measured at 300 K (panel b)) and the USG measured at 150 K (panel c)). The evaluated $C_p$ is represented as a red solid line, its harmonic contribution as yellow dashed line and the TLS contribution as dotted violet lines. The Debye levels found with the fitting procedure are depicted with black solid lines, while the values found with Brillouin spectroscopy with dashed black lines.} \label{Figs/Analysis: Cp_fit3}
\end{figure*}

The total specific heat includes a linear term describing the effect of two level states, resulting in eq. (3) of the main text, that we report here for clarity:

\begin{widetext}
\begin{equation}
\label{Eq: cp fit}
            C_p(T) = \frac{3}{M_{at}}\int d\nu g(\nu)\frac{\partial n(T,\nu)}{\partial T} h\nu + C_{TLS}  T\\ 
     = \frac{3}{M_{at}} \left (
    \int_{0}^{\nu_0} d\nu g_{LF}(\nu)\frac{\partial n(T,\nu)}{\partial T}h\nu + \int_{\nu_0}^{\infty} d\nu g_{exp}(\nu)\frac{\partial n(T,\nu)}{\partial T}h\nu \right ) + C_{TLS} T  .
\end{equation}
\end{widetext}

In this expression $M_{at}$ is the average atomic mass, $k_B$ the Boltzmann constant, $g_{LF}(\nu)$ is modelled as in eq.~\eqref{eq:parabola} and $g_{exp}(\nu)$ is the measured DOS.
The frequency $\nu_0$, the Debye frequency $\nu_D$ and $C_{TLS}$ are treated as fitting parameters to the experimental specific heat. It is worth noting that $\nu_0$, $\nu_D$ and $A_{ex}$ are linked by eq.~\eqref{eq:constraint} so that we can choose $A_{ex}$ in place of $\nu_0$ as free parameter, as done in the main text. The standard tunnelling model~\cite{Ramosbook} gives a relationship between $C_{TLS}$ and the density of TLS per atom and per unit frequency, $n_{TLS}$, that we assume constant in energy:
\begin{equation}\label{eq:nTLS}
    C_{TLS} = \frac{\pi^2}{6 M_{at}} \frac{n_{TLS}}{h} k_B^2,
\end{equation}
where $h$ is the Planck constant.
The comparison between the evaluated and the measured specific heats is reported in Figures (\ref{Figs/Analysis: Cp_fit3 usg}-\ref{Figs/Analysis: Cp_fit3 150}). In table~\ref{tab:my-table} we summarize the obtained values for $\nu_0$, $3/\nu_D^3$ and $n_{TLS}$. The parameters derived using the DOS of the USG at room temperature are compatible within uncertainty with those found using the DOS at 150 K. In the main text we have used the latter, because it is closer to the harmonic DOS, being measured at a lower temperature.
\begin{table}[htb]
\centering
\caption{Best fitting parameters found for the three cases.}
\label{tab:my-table}
\begin{tabular}{c|c|c|c}
  \hline
  \hline
 &
  \begin{tabular}[c]{@{}c@{}}$\nu_0$\\ {[}THz{]}\end{tabular} &
  \begin{tabular}[c]{@{}c@{}}$3/\nu_D^3$\\ {[}THz$^{-3}${]}\end{tabular} &
  \begin{tabular}[c]{@{}c@{}}$n_{TLS}$\\ {[}THz$^{-1}$atoms$^{-1}${]}\end{tabular}  \\ \hline
USG 300 K  & 0.20 $\pm$ 0.01 & 0.036 $\pm$ 0.004 & (4 $\pm$ 2)$\times$10$^{-5}$  \\ \hline
OG - 300 K  & 0.279 $\pm$ 0.005 & 0.039 $\pm$ 0.002 & (20 $\pm$ 1)$\times$10$^{-5}$  \\ \hline
USG - 150 K & 0.19 $\pm$ 0.01 & 0.038 $\pm$ 0.003 & (4 $\pm$ 2)$\times$10$^{-5}$  \\
  \hline
  \hline
\end{tabular}
\end{table}

Figures (\ref{Figs/Analysis: rdos usg}-\ref{Figs/Analysis: rdos 150}) report the comparison between the measured DOS and its low temperature limit estimated from the fit to the specific heat, for all the samples. In the OG the parabolic extrapolation of the reduced DOS to low frequencies reaches a Debye level that coincides with the low T estimate from Brillouin scattering measurements, as shown in Figure (\ref{Figs/Analysis: Cp_fit3 og}), while the Debye level for the USG glass is significantly higher than the low T estimate from the elastic moduli, revealing the presence of elastic anisotropy, as discussed in the main text. 

The specific heat at low temperatures can be approximated by neglecting the high frequency part of the DOS, obtaining:

\begin{widetext}
\begin{equation}
    C_p(T) \sim C_{TLS} T + \frac{3}{M_{at}} \int_{0}^{\infty} d\nu g_{LF}(\nu)\frac{\partial n(T,\nu)}{\partial T}h\nu 
    = C_{TLS} T + C_D T^3 + C_{ex} T^5 ,
\end{equation}
\end{widetext}
where $C_D$ and $C_{ex}$ are related to $\nu_D$ and $A_{ex}$ by the following relations:
\begin{align}
    C_D =& \frac{12 \pi^4}{5} \frac{k_B}{M_{at}} \left( \frac{k_B}{h}\right)^3 \frac{1}{\nu_D^3} ,  \\
    C_{ex} =& \frac{16 \pi^6}{7} \frac{k_B}{M_{at}} \left( \frac{k_B}{h}\right)^5 A_{ex} .
\end{align}
We use these relationships to estimate $\nu_D$ and $A_{ex}$ for the crystal from a fit to its low temperature specific heat, as discussed in the main text.

\section{Estimate of the low temperature Debye levels from Brillouin light scattering and comparison with the isotropic USG}\label{App:sound vel}
\begin{figure}[ht]
    \centering

    \subfloat[]{%
      \includegraphics[width=1\linewidth]{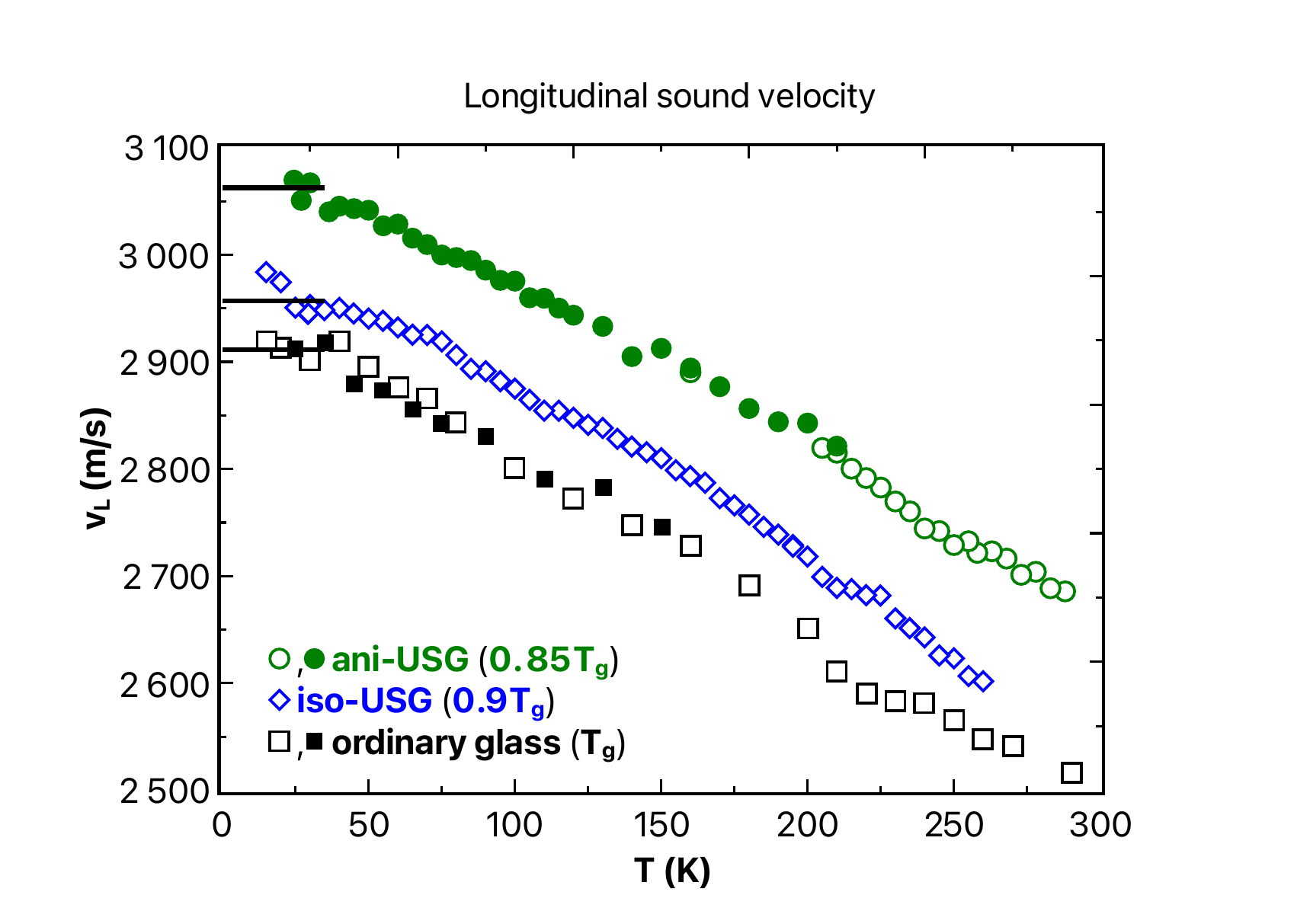}
    }

    \subfloat[]{%
      \hspace*{-7mm}\includegraphics[width=0.9\linewidth]{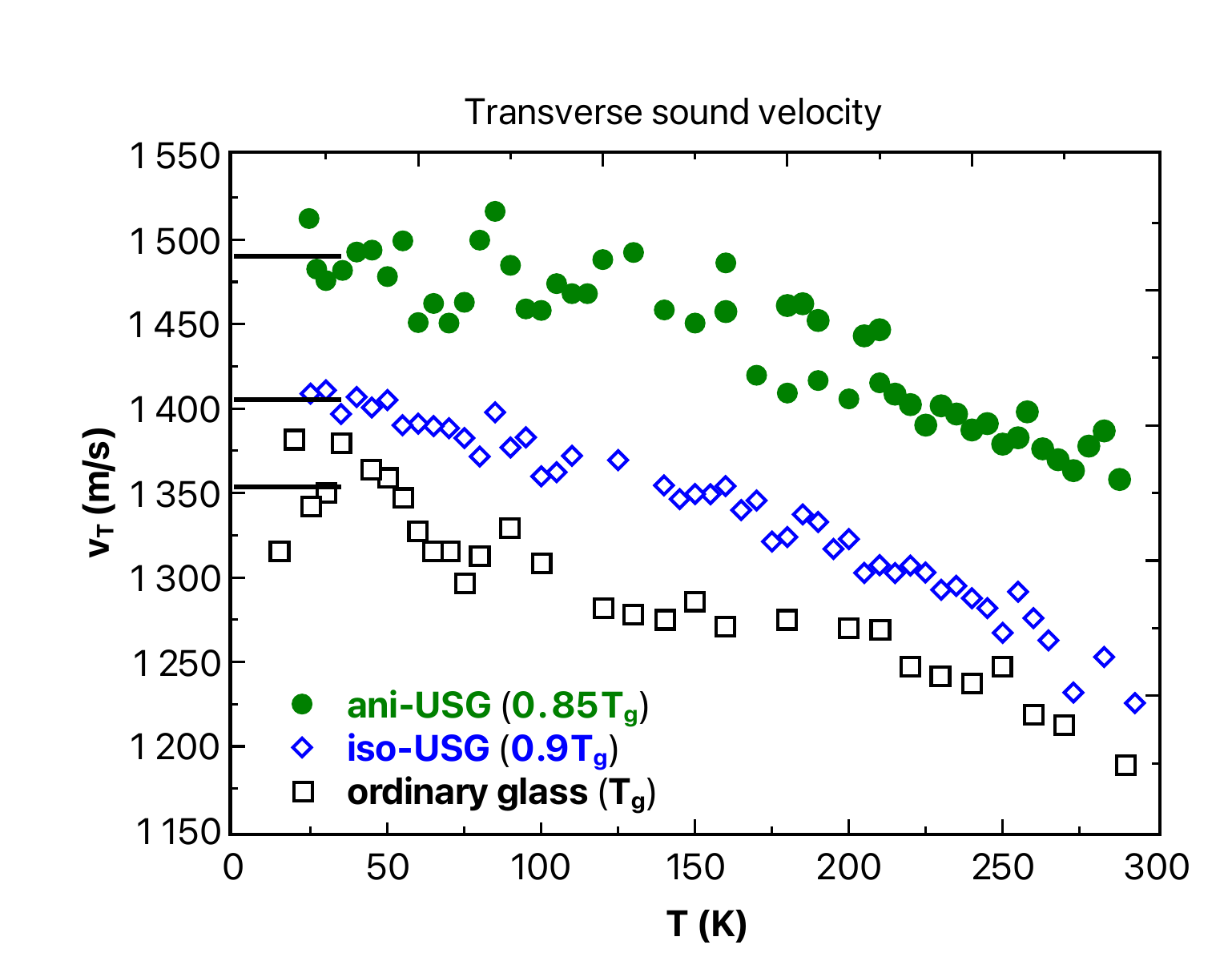}
    }

    \caption{The longitudinal (panel a)) and the transverse (panel b)) sound velocities measured with Brillouin light scattering~\cite{Moratalla2023}. The green circles represent the data for the ultra-stable glass deposited at 0.85$T_g$ (ani-USG), the blue diamonds the USG deposited at 0.9$T_g$ (iso-USG) and the black squares the ordinary glass. The horizontal black lines are the velocity values obtained by averaging temperatures below 35 K.}
    \label{Figs/sound vel}
\end{figure}

The Debye frequencies were estimated from the longitudinal and transverse sound velocities measured by Brillouin Light Scattering~\cite{Moratalla2023}, as reported in Fig. \ref{Figs/sound vel}. The figure shows data for the ordinary glass, the ultra-stable glass deposited at 0.85$T_g$ (ani-USG), and the ultra-stable glass deposited at 0.9$T_g$ (iso-USG), respectively, as black squares, green circles, and blue diamonds.

The temperature dependence is not linear across the entire range, particularly for the transverse velocity, which dominates the determination of the Debye level. For this reason, we assumed that both $v_L$ and $v_T$ remain constant below $T \le 35$ K.
The Debye frequencies obtained from the averaged low-temperature sound velocities are reported in table \ref{T:velsound}.

\begin{table}[ht]
\centering
\caption{Debye frequencies for the OG, the iso-USG and the ani-USG, obtained from the averaged low-temperature sound velocities.}
\label{T:velsound}
\begin{tabular}{c|c}
  \hline
  \hline
 &
  \begin{tabular}[c]{@{}c@{}}$\nu_D^{BLS}$\\ {[}THz{]}\end{tabular} \\ \hline
Ordinary glass  & 4.21 $\pm$ 0.07 \\ 
Isotropic USG  & 4.38 $\pm$ 0.07  \\ 
Anisotropic USG & 4.64 $\pm$ 0.07  \\
  \hline
  \hline
\end{tabular}
\end{table}

These values are compared with those derived from the specific-heat analysis discussed in the main text, which yield $\nu_D = (4.27 \pm 0.06)$ THz for the ordinary glass and $(4.29 \pm 0.08)$ THz for the ultra-stable glass. The comparison gives $\nu_D - \nu_D^{BLS} = (0.06 \pm 0.09)$ THz for the ordinary glass and $(-0.10 \pm 0.10)$ THz for the isotropic ultra-stable glass, indicating that in both cases the two estimates are consistent within uncertainty. For the anisotropic ultra-stable glass, the difference is $(-0.35 \pm 0.10)$ THz, confirming its elastic anisotropy, since $\nu_D^{BLS}$, obtained from sound waves propagating parallel to the substrate, is higher than the average value that gives rise to the proper Debye frequency sensed by the specific heat.

\section{The soft potential model}\label{App:SPM}
The soft potential model (SPM) is based on the assumption of the presence of a distribution of soft anharmonic potentials that can be written in appropriate units in the form~\cite{Ramos1993}:
\begin{equation}
    V(x) =  W (D_1 x + D_2 x^2 + x^4) ,
\end{equation}
where $x$ is a dimensionless displacement, $D_1$ and $D_2$ are parameters that vary in different positions within the glass, and $W$ is a characteristic energy scale that can be assumed constant for all the soft modes.
The model assumes a random distribution of the parameters $D_1$ and $D_2$ around the origin of the $D_1 - D_2$ plane with a density $P(D_1,D_2)=P_s$. The model further assumes that the interaction between the soft modes and the sound waves is linear in both the displacement $x$ of the mode and in the strain field of the sound wave. The model has, consequently, a total of three independent parameters: $P_s$, $W$ and $\Lambda$, the coefficient of the interaction between soft modes and sound waves. It is worth noting that this third parameter is required to describe the sound attenuation and the thermal conductivity but does not enter in the density of states and specific heat estimates.
The soft potential model is in agreement with the predictions of the tunnelling model at low temperatures. In particular, the linear growth of the specific heat with temperature is described by a coefficient $C_{TLS}$ given by~\cite{Ramos1993}:
\begin{equation}\label{eq:SPMTLS}
    C_{TLS} \sim \frac{10}{M_{at}} \frac{P_s}{W} k_B^2 .
\end{equation}
From this expression we can estimate the ratio $P_s/W$ for the OG and the USG glasses.
The model predicts the presence of tunnelling states with a density proportional to $P_s/W$ at frequencies below $W/h$. At higher frequencies, instead, the model predicts the presence of harmonic vibrations. The density of the soft quasi-localized modes that are in excess of the sound waves grows as the fourth power of frequency, as~\cite{Buchenau2007}:
\begin{equation}\label{eq:SPMdos}
    g_s(\nu) = \frac{1}{24} \frac{P_s h}{W^5} (h \nu)^4 .
\end{equation}
\begin{table}[ht]
\centering
\caption{Soft potential model parameters for the USG and the OG, in the hypothesis that the departure from the Debye level in the harmonic DOS is entirely due to QLMs.}
\label{tab:SPM}
\begin{tabular}{c|c|c}
  \hline
  \hline
 &
  \begin{tabular}[c]{@{}c@{}}$P_s$\\ {[}state/atoms{]}$\times$10$^{-6}$\end{tabular} &
  \begin{tabular}[c]{@{}c@{}}$W/h$\\ {[}GHz{]}\end{tabular} \\
  \hline
USG & 0.17 $\pm$ 0.09   & 25 $\pm$ 3  \\ \hline
OG  & 1.25 $\pm$ 0.07   & 38.1 $\pm$ 0.6 \\ 
  \hline
  \hline
\end{tabular}
\end{table}
From the values of $n_{TLS}$ and of $A_{ex}$ we estimate the parameters $P_s$ and $W$ for the OG and USG using equations~\eqref{eq:SPMTLS} and~\eqref{eq:SPMdos}. The SPM parameters are reported in table~\ref{tab:SPM}.

\begin{figure}[ht]
    \centering
        \includegraphics[width=0.4\textwidth]{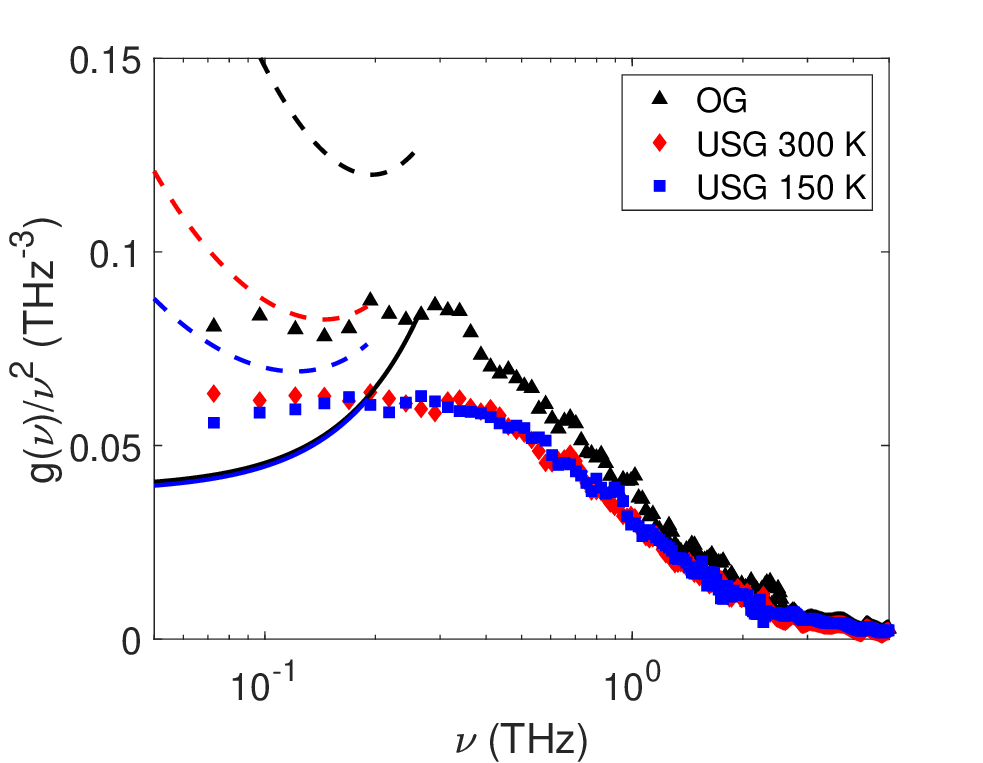}
    \caption{SPM prediction for the relaxational contribution to the DOS, in the hypothesis that the departure from the Debye level in the harmonic DOS is entirely due to QLMs. Reduced DOS of the OG and USG with symbols as in the legend. The continuous lines are our estimate of the low frequency harmonic DOS (black line for the OG, blue for the USG). The dashed lines are the SPM estimates of the low-frequency part of the DOS, including both the harmonic contribution of eq.~\eqref{eq:SPMdos} and the relaxational part of eq.~\eqref{eq:SPMrel}. The blue line is calculated at 150 K with the SPM parameters of the USG, the red line at 300 K for the USG and the black line at 300 K for the OG.}
    \label{Figs/DOSrel}
\end{figure}

The SPM predicts a relaxational contribution to the vibrational DOS, approximated by eq~\eqref{eq:SPMrel} of the discussion section. The reduced DOS calculated by summing the contributions of eq.~\eqref{eq:SPMdos} and eq.~\eqref{eq:SPMrel} are reported as dashed lines in Figure~\ref{Figs/DOSrel}, showing that the SPM overestimates the relaxational component by approximately a factor of two. This is most probably due to an overestimate of the QLMs, since the model neglects the importance of sound attenuation in the departure of the DOS from the Debye level. If we fix the SPM parameters in such a way that equation~\eqref{eq:SPMrel} can properly describe the anharmonic component of the DOS, we obtain a more reliable estimate of the QLM density, as discussed in the main text.

\section{The heterogeneous-elasticity theory}\label{App:HET}

\begin{figure*}[ht]
    \centering
\subfloat[\label{fig:7deg:a}]{%
  \includegraphics[width=0.4\textwidth]{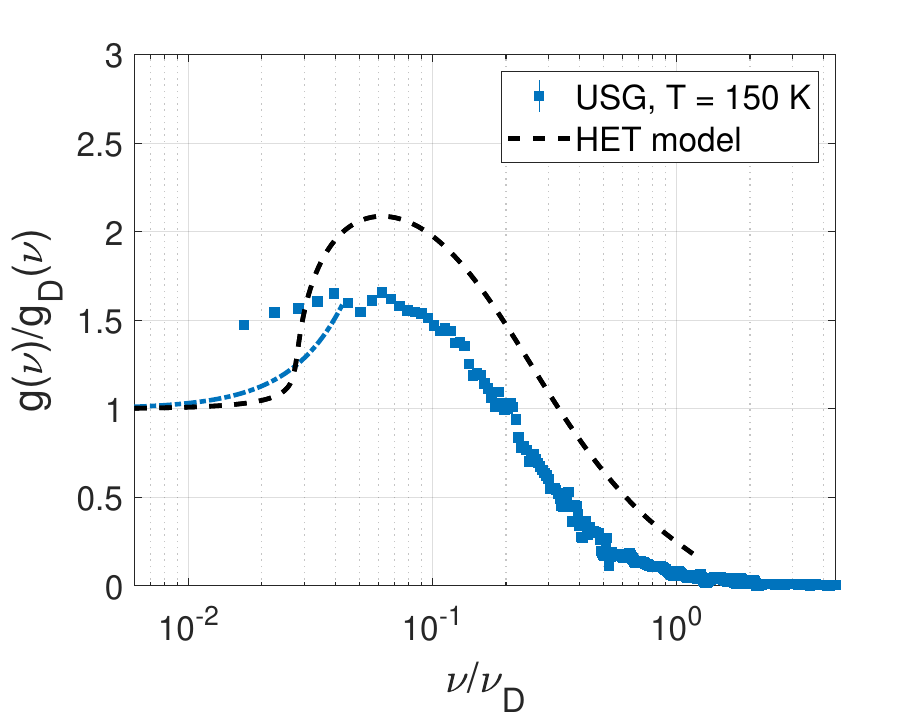}%
}\hspace*{\fill}%
\subfloat[\label{fig:7deg:b}]{%
  \includegraphics[width=0.4\textwidth]{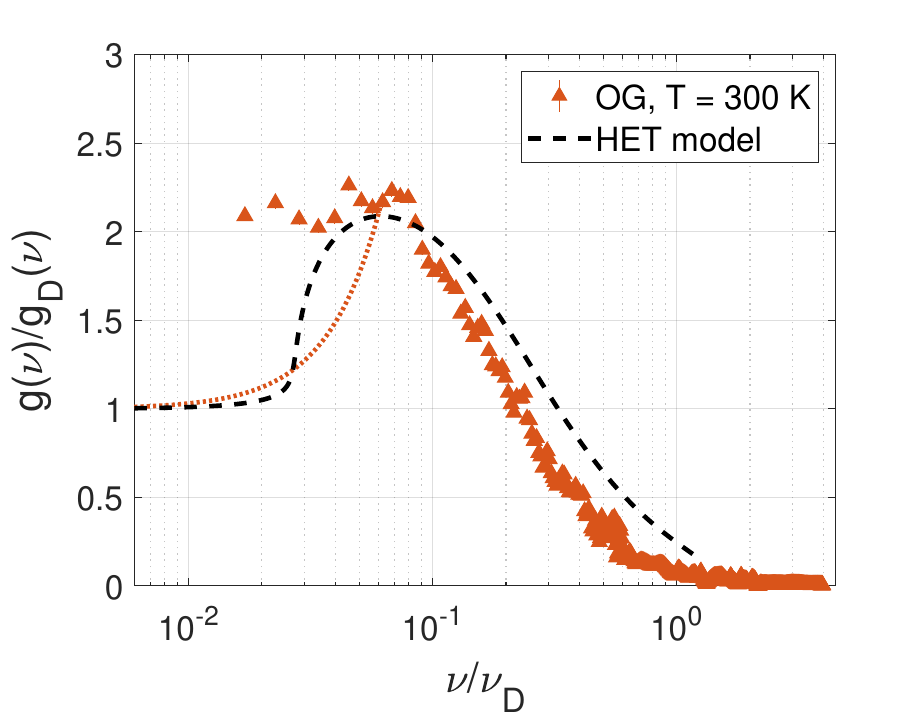}%
}
\caption{The reduced density of states in Debye units. The DOS $g(E)$ normalised to the Debye contribution $g_D(\nu) = 3\nu^2/\nu_D^3$ compared to the curve obtained applying the HET, normalised by the DL found from $k_D$, $v_{L,0}$ and $v_{T,0}$. Panel a): USG. Panel b): OG.} \label{Figs: model HET}
\end{figure*}
We present here the heterogeneous-elasticity theory~\cite{articlemodel,PhysRevLett.100.137402} derived in the self consistent Born approximation. This approach is equivalent to the more recent version of the theory~\cite{SCHIRMACHER2015133,PhysRevB.104.134106} in the limit of small disorder. The theory assumes that the elastic constants spatially fluctuate with a correlation length $\xi$. The spatially fluctuating shear modulus $G(\textbf{r})=G_0+\Delta G(\textbf{r})$ has fluctuations $\Delta G(\textbf{r})$ correlated according to a correlation function $C_G(\textbf{r}) = \langle\Delta G(\textbf{r}+\textbf{r}_0)\Delta G(\textbf{r}_0)\rangle$. The $C_G(\textbf{r})$ and its Fourier transform are
\begin{equation}
    \begin{aligned}
        C_G(\textbf{r}) & = \langle\Delta G^2\rangle \text{e}^{-r/\xi}\\
        C_G(\textbf{k}) & = \langle \Delta G^2\rangle (8\pi/\xi)[k^2+\xi^{-2}]^{-2} \ .
    \end{aligned}
\end{equation}
The corresponding mean-field equation for the low-wavenumber self energy $\Sigma(\omega)=\Sigma(q=0,\omega)$ in the self-consistent Born approximation is:  

\begin{equation}
    \begin{aligned}
            \Sigma(\omega) = & \frac{\gamma}{2\varphi_3\langle \Delta G^2\rangle}\int\displaylimits_{|\textbf{k}|\text{\textless}k_D}\left(\frac{d\textbf{k}}{2\pi}\right)^3\Big(C_G(k) \\
             & \times [\chi_L(k,\omega)+\chi_T(k,\omega)] \Big)
    \end{aligned}
\end{equation}

where $\chi_L(k,\omega)$ and $\chi_T(k,\omega)$ are the longitudinal and transverse dynamical susceptibilities:
\begin{equation}
    \begin{aligned}
        \chi_L(k,\omega) & = k^2[-\omega^2 + k^2(v_{L,0}^2 - 2\Sigma(\omega))]^{-1}\\
        \chi_T(k,\omega) & = k^2[-\omega^2 + k^2(v_{T,0}^2 - \Sigma(\omega))]^{-1} \ .
    \end{aligned}
\end{equation}
The prefactor $\gamma=\langle \Delta G^2\rangle\varphi_3/v_{T,0}^4$ is the \textit{disorder parameter}, $v_{L,0}$ and $v_{T,0}$ are the longitudinal and transverse sound velocities, and $\varphi_3=\int\displaylimits_{|\textbf{k}|\text{\textless}k_D}(d\textbf{k}/2\pi)^3C_G(\textbf{k})$ is a normalization constant. The integral is computed up to the Debye wavevector $k_D$. \\

The density of vibrational states is evaluated as:
\begin{equation}
    g(\omega) = \frac{2\omega}{3\pi}\int\displaylimits_{|\textbf{k}|\text{\textless}k_D}\left(\frac{d\textbf{k}}{2\pi}\right)^3\frac{1}{k^2}\text{Im}\{\chi_L(k,\omega) + 2\chi_T(k,\omega)\} \ .
    \label{Eq: model g(w)}
\end{equation}
The Debye wavevector $k_D$ and the zero temperature longitudinal and transverse velocities ($v_{L,0}$, $v_{T,0}$) are needed for both glasses, in order to evaluate the DOS of the model. The Debye wavevector $k_D$ is calculated using their mass densities ($\rho_{USG} = 1.066$ g/cm$^3$, $\rho_{OG} = 1.05$ g/cm$^3$) and the mean atomic weight of the TPD molecule ($\simeq 7.18$ g/mol). The $v_{L,0}$ and $v_{T,0}$ are derived from Brillouin scattering measurements~\cite{Moratalla2023} performed on equivalent samples. However, to better describe the position of the boson peak in the specific heat (see Fig. 4 on the right of \cite{festi}), we changed the extrapolated values of longitudinal and transverse velocities of a 1-2\% factor.  The parameter values are summarised in the following table for the ultrastable and the conventional glasses:
\begin{table}[ht]
\centering
\caption{HET parameters for the USG and OG glasses.}
\label{tab:het}
\begin{tabular}{c|cc|c|c}
  \hline
  \hline
 \ & $v_{L,0}$ & $v_{T,0}$ & $k_D$ & $\gamma$ \\
 \ & {[}m/s{]} & {[}m/s{]} & {[}\AA$^{-1}${]} & \\ \hline
USG  & 3067  & 1493  & 1.74 & 0.404\\
OG   & 3022  & 1392  & 1.73 & 0.413\\ 
  \hline
  \hline
\end{tabular}
\end{table}
The models of the reduced DOS are shown in Supplementary Figure \ref{Figs: model HET} for both glasses. The position of both boson peaks is well described, as the height of the BP of the ordinary glass. The height of the BP for the USG is instead overestimated. The HET predicts a BP intensity twice the value of the Debye level, as seen for the OG. The USG is characterised by unusually low excess of vibrational modes, as highlighted in the main text.  

It is worth noting that these estimates are based on the Brillouin values of the sound velocities, as indicated in table~\ref{tab:het}. These values are appropriate for the OG but are overestimated for the USG, because of the elastic anisotropy discussed before. However, the difference between the HET curves in Figure~\ref{Figs: model HET} for the USG and the OG is very weakly affected by the sound velocity values and depends only on the parameter $\gamma$. The sound velocities in the HET model affect the value of the Debye level but not the excess over the Debye.

\bibliography{bibliog}

\end{document}